\documentclass[11pt]{article}

\usepackage[letterpaper,margin=1in]{geometry}
\usepackage[T1]{fontenc}
\usepackage{lmodern}
\usepackage{microtype}
\usepackage{amsmath,amssymb,amsthm,mathtools}
\usepackage{algorithm}
\usepackage{float}
\usepackage{algpseudocode}
\usepackage{booktabs}
\usepackage{array}
\usepackage{enumitem}
\usepackage{needspace}
\usepackage{xcolor}
\usepackage{tikz}
\usepackage{pgfplots}
\usepackage{appendix}
\pgfplotsset{compat=1.18}
\usetikzlibrary{arrows.meta}
\usepackage{hyperref}
\usepackage[nameinlink,noabbrev]{cleveref}

\makeatletter
\providecommand*{\theHALG@line}{}
\renewcommand*{\theHALG@line}{\thealgorithm.\arabic{ALG@line}}
\makeatother

\theoremstyle{plain}
\newtheorem{theorem}{Theorem}[section]
\newtheorem{lemma}[theorem]{Lemma}
\newtheorem{proposition}[theorem]{Proposition}
\newtheorem{corollary}[theorem]{Corollary}

\theoremstyle{definition}

\newtheorem{remark}[theorem]{Remark}
\newtheorem{example}[theorem]{Example}

\newcommand{\ket}[1]{\lvert #1\rangle}
\newcommand{\bra}[1]{\langle #1\rvert}

\newcommand{\wt}{\operatorname{wt}}
\newcommand{\C}[2]{\mathsf{C}_{#1}(#2)}

\newcommand{\makefrontmatter}{%
  \begingroup
  \renewcommand{\thefootnote}{\fnsymbol{footnote}}%
  \begin{center}
    {\Large\bfseries An Efficient Quantum Algorithm for Preparing Uniform
    Superpositions over Arbitrary Integer Intervals \par}
    \vspace{1.1em}

    {\large
    Alok Shukla$^1$\footnotemark[1] and Prakash Vedula$^2$\par}

    \vspace{0.75em}
    {\normalsize
    $^1$School of Arts and Sciences, Ahmedabad University, Ahmedabad, India\\
    $^1$\texttt{alok.shukla@ahduni.edu.in}\\[0.45em]
    $^2$School of Aerospace and Mechanical Engineering,
    University of Oklahoma, Norman, USA\\
    $^2$\texttt{pvedula@ou.edu}\par}

    \vspace{0.65em}
    {\normalsize \today\par}
  \end{center}

  \footnotetext[1]{Corresponding author%:
  }
  \endgroup
  \vspace{0.8em}
}

\renewenvironment{abstract}{%
  \begin{center}\bfseries Abstract\end{center}
  \small\begin{quote}\ignorespaces
}{\end{quote}\normalsize\vspace{0.35em}}

\begin{document}
\makefrontmatter
\begin{abstract}
We consider exact preparation of the uniform quantum state
supported on an arbitrary nonempty interval of integer labels
in an $n$-qubit register. For every
$0\leq a<a+M\leq2^n$, we give an ancilla-free circuit using
only one-qubit gates and singly controlled one-qubit gates.
In the logical gate model specified in the paper, the direct
construction has exact worst-case gate count $5n-9$ for
$n\geq3$; its exact maxima for $n=1$ and $n=2$ are $1$ and
$3$, respectively. A worst-case lower bound of $\Omega(n)$
establishes asymptotic optimality in the register width.
In addition, the gate complexity admits an instance-dependent
bound $O(\log_2 M+\epsilon)$, where $\epsilon$ is an
offset-dependent parameter defined in terms of binary
Hamming weights.
We also derive a reflected construction and a classical
procedure that selects the lower-count exact circuit.
The direct construction admits an $O(n)$ logical-depth
upper bound. We also treat bitwise-XOR images of
intervals and arithmetic progressions with power-of-two
spacing.
\end{abstract}

% \noindent{\footnotesize
% \textit{2020 Mathematics Subject Classification.} Primary 81P68; Secondary 68Q12.\\
% \textit{Key words and phrases.} Quantum state preparation, uniform superpositions, integer intervals, controlled rotations, quantum circuits, circuit complexity.}
%\vspace{0.8em}

% \clearpage

\section{Introduction}

Quantum computing has emerged as a powerful computational paradigm, motivating the development of quantum algorithms to address computational problems that can be intractable for classical methods. Over the past decades, a substantial body of quantum algorithms has been developed, with provable advantages for specific structured computational tasks. Representative examples include the Deutsch--Jozsa algorithm \cite{deutsch1992rapid}, the Bernstein--Vazirani algorithm \cite{bernstein1993quantum} and its probabilistic generalization \cite{shukla2023generalizationBV}, Simon's algorithm \cite{simon1997power}, and quantum search \cite{grover1996fast,shukla2025efficientGrover}. Important algorithmic primitives include the quantum Fourier transform and quantum phase estimation \cite{nielsen2002quantum,kitaev1995quantum,dobvsivcek2007arbitrary,shukla2025towardsQPE}, as well as quantum amplitude estimation \cite{brassard2002quantum,giurgica2022low,shukla2025modularQAE}. Shor's polynomial-time algorithms for integer factorization and discrete logarithms provide another prominent example of quantum speedup \cite{shor1999polynomial,shukla2025modularShor}.

Beyond these foundational algorithms, quantum computing has increasingly been explored as a computational tool for scientific and engineering applications. These include trajectory optimization \cite{shukla2019trajectory}, the solution of large-scale linear systems \cite{harrow2009quantum}, linear and nonlinear differential equations \cite{childs2020quantum,berry2014high,shukla2023hybriddiffeq}, digital image and signal processing, and spectral analysis \cite{shukla2023hybridimageprocessing,shukla2023quantumDSP,rohida2024hybridPolar,rohida2025quantumEdgeDetect}. Other areas of investigation include quantum optimal control \cite{sandesara2024quantumOptimalcontrol}, quantum machine learning \cite{biamonte2017quantum}, and numerical integration \cite{shimada2020quantum,abrams1999fast,shukla2025efficientPartialSum}. These applications often involve processing structured classical data, such as discretized functions, images, signals, coefficient arrays, or training data, within a quantum computational framework.
In order to realize the full potential of such algorithms, efficient preparation of their input quantum states is often required. 
To assess end-to-end quantum advantage in practical
applications, one must account not only for the cost of the
main quantum routines but also for state preparation and
measurement. The existence of an efficient quantum routine
does not, by itself, guarantee an end-to-end quantum advantage
if the costs of preparing the input state or obtaining the
classical output through measurement become prohibitively high.
This issue is particularly significant in application-oriented settings, including differential equation solvers, quantum image and signal processing, quantum optimal control, and quantum machine learning, where the input data may have a nontrivial structure and must be encoded repeatedly or at different stages of a computation. Therefore, efficient state preparation should be regarded as an integral part of quantum algorithm design, and its resource requirements must be included in any meaningful assessment of computational advantage. This motivates the development of exact and resource-efficient procedures for preparing structured quantum states, including uniform superpositions over prescribed sets of computational-basis states.

Uniform superpositions over structured subsets of computational-basis
states are used in many important quantum algorithms, including quantum search, amplitude amplification, numerical integration, sampling, and other algorithmic
routines \cite{grover1997quantum, shukla2025efficientGrover, shukla2025efficientPartialSum}. In this work, we propose an efficient algorithm for the preparation of the uniform quantum state on an
arbitrary nonempty integer interval. For an $n$-qubit register and
integers $0\leq a<a+M\leq2^n$, the target is
\begin{equation}\label{eq:target}
  \ket{\psi_{a,M}^{(n)}}=\frac{1}{\sqrt M}
  \sum_{x=a}^{a+M-1}\ket{x}.
\end{equation}
Throughout this work we will use the half-open interval notation $[a,a+M)$.

An important special case is $a=0$, which gives the uniform state on
the initial segment $[0,M)$. We call this the \emph{prefix state}.
In Ref.~\cite{shukla2024efficient}, we introduced an exact,
ancilla-free algorithm for preparing this state for arbitrary
$1\leq M\leq 2^n$, including values of $M$ that are not powers of
two. By bitwise complementation, the same construction also prepares
the corresponding \emph{suffix state}, supported on the terminal
segment $[2^n-M,2^n)$. We refer to this earlier method as the
\emph{uniform-superposition preparation algorithm}. The corresponding
\texttt{UniformSuperpositionGate} implementations in Qiskit and Cirq
also use this algorithm
\cite{QiskitUniformSuperpositionGate,CirqUniformSuperpositionGate}.

We give exact, ancilla-free circuits using only one-qubit gates
and singly controlled one-qubit gates for the preparation of
the interval state $\ket{\psi_{a,M}^{(n)}}$. In the logical
gate model defined below, the direct construction has exact
worst-case gate count $5n-9$ for $n\geq3$, while its exact
maxima for $n=1$ and $n=2$ are $1$ and $3$, respectively.
A worst-case lower bound of $\Omega(n)$ establishes
asymptotic optimality in register width. 
The worst-case bound is expressed in terms of the register width
$n$, but the cost of an individual interval can be substantially
smaller. We establish an additional instance-dependent bound
\[
  G_{\mathrm{dir}}(a,M)
  \leq 4\lfloor\log_2 M\rfloor+\epsilon(a,M),
  \qquad M>1,
\]
where $\epsilon(a,M)$ is defined in \Cref{sec:complexity}.
It is a sum of binary Hamming weights associated with the common
high-order prefix and, in the generic case, the final branch-dependent
bit masks. Thus, the local preparation cost grows logarithmically
with the interval length, while $\epsilon$ accounts for the
additional cost associated with the position and binary structure
of the interval within the register. In particular, intervals with
a small value of $\epsilon$ admit logarithmic gate complexity even
when embedded in a register of much larger width.

In this work, we also provide another method for interval state preparation, referred to as the reflected
construction of the interval, and a classical compiler that selects the lower-count
circuit for each instance after comparing the cost of the direct and the reflected construction. Furthermore, we also prepare states with special structures such as in the form of arithmetic progressions with power-of-two spacing, by using the bitwise-XOR images of intervals.

The paper is organized as follows. Related
work is reviewed in \Cref{sec:related_work}.  The direct interval state
construction, including the common dyadic-block partition,
the paired-angle implementation, and the XOR conversion is described in \Cref{sec:interval-prep}. Next, 
\Cref{sec:direct-correctness} establishes the correctness of the direct
construction.  \Cref{sec:reflected-construction} presents the reflected construction.
\Cref{sec:xor-progressions} develops the XOR-image and arithmetic-progression
extensions. \Cref{sec:complexity} analyzes gate complexity and logical
depth and describes adaptive circuit selection. \Cref{sec:examples}
provides worked examples. \Cref{sec:conclusion} presents the
conclusion.

The appendices provide technical details of the reflected
construction, additional complexity proofs and interval
identities, and extensions to nonuniform block-weighted
states and other interval constructions.

\section{Related work and comparison}
\label{sec:related_work}

General pure-state synthesis uses multiplexed rotations or universal
few-qubit decompositions and has exponential worst-case cost in the
number of qubits for unrestricted target states
\cite{mottonen2004transformation,ShendeBullockMarkov2006,PleschBrukner2011}.
Structured methods can improve on this when the amplitudes or support
have additional structure. Grover-Rudolph preparation uses conditional
rotations determined by cumulative probability masses
\cite{GroverRudolph2002}; quantum eigenvalue-transformation methods
provide a broader framework for preparing function-defined amplitude
profiles \cite{McArdleGilyenBerta2026}.

The uniform-superposition preparation algorithm of
Ref.~\cite{shukla2024efficient} gives an exact, ancilla-free
preparation of the prefix state on $[0,M)$, with gate complexity and
depth $O(\log M)$, using $\lceil\log_2 M\rceil$ qubits and no
multi-controlled gates. By bitwise complementation, it also gives
the corresponding suffix state. We use the prefix-state circuit
when the target interval is a prefix of an aligned interval of
power-of-two length. The present work treats arbitrary offsets and
proves an explicit worst-case bound in the full register width.

% For the preparation of the prefix state on $[0,M)$, reference~\cite{babbush2018encoding} provides an amplitude amplification based approach 
% using $\log_2(M)$ ancillary qubits.
% In contrast, our construction is ancilla-free.

For prefix-state preparation, a method is described in Fig.~12 of the reference~\cite{babbush2018encoding}. It uses 
controlled amplitude-amplification construction for preparing
the uniform superposition over $M=2^rL$ labels, where $L$
is odd. Their circuit has total width
   $ r+2\log_2L+O(1)$
qubits and a $T$-gate count
   $ 2r+10\log_2L+O(\log(1/\varepsilon))$,
where $\varepsilon$ denotes the rotation-synthesis
precision parameter. Since the output index register requires
$r+\lceil\log_2L\rceil$ qubits, the construction uses
$O(\log L)$ additional workspace qubits. In contrast, our construction is ancilla-free.

Other approaches address either arbitrary support sets or one interval
through a separate state preparation and translation. The
\texttt{SQUARE} construction in PyEncode prepares the prefix state
using the uniform-superposition preparation algorithm of
Ref.~\cite{shukla2024efficient} and then applies a constant adder
based on the quantum Fourier transform
(QFT); its general cost is quadratic in the register width, with
linear-cost special cases
\cite{Suresh2026PyEncode,Draper2000}. Boolean-function and
decision-diagram methods cover broader support families, with costs
depending on the representation of the support and, in some cases, an
ancillary qubit \cite{MozafariRienerSoekenDeMicheli2021,MozafariDeMicheliYang2022}.
The present construction uses the binary structure of the two interval
endpoints to obtain a uniform linear bound for every translated
interval.

\begin{table}[H]
\centering
\caption{Comparison with related state-preparation methods. Complexity bounds are those reported or derived from the cited constructions; the gate models and ancillary-qubit conventions are not identical.}

\label{tab:related-work}
\small
\begin{tabular}{@{}>{\raggedright\arraybackslash}p{3.5cm}>{\raggedright\arraybackslash}p{4cm}>{\raggedright\arraybackslash}p{7.5cm}@{}}
\toprule
Method & Target family & Construction and reported complexity \\
\midrule
Shukla and Vedula~\cite{shukla2024efficient} & Initial interval $[0,M)$ & Exact prefix-state preparation with $O(\log M)$ gate complexity and depth, without ancillary qubits or multi-controlled gates. \\ \\
PyEncode \texttt{SQUARE} \cite{Suresh2026PyEncode} & One arbitrary interval & $O(n^2)$ for general offsets; $O(n)$ when the start is zero or the interval is an aligned block of power-of-two length. \\ \\
Boolean-function uniform state preparation \cite{MozafariRienerSoekenDeMicheli2021} & General uniform supports & Exact construction based on binary decision diagrams, without ancillary qubits; exponential worst-case CNOT count. \\ \\
Decision-diagram method \cite{MozafariDeMicheliYang2022}
& General states represented by reduced decision diagrams
& $O(nP)$ CNOT gates, where $P$ is the number of
decision-diagram paths; one ancillary qubit.  \\ \\
This work
& Arbitrary interval $[a,a+M)$
&  Ancilla-free. Exact worst-case logical gate count
$5n-9$ for the specified direct construction when
$n\geq3$; instance-dependent complexity
$O(\log_2 M+\epsilon(a,M))$. Also provides a reflected construction
and adaptive selection by complete logical gate count. \\ 
\bottomrule
\end{tabular}
\end{table}

\begin{remark}
    The parameter $\epsilon(a,M)$ in the row corresponding to this work
is defined in Proposition~\ref{prop:instance-dependent-bound}.
It captures the offset-dependent contribution to the gate count
through the Hamming weights of the common high-order prefix and,
for a generic midpoint split, the final branch-dependent masks.
\end{remark}

\section{Preparation of a uniform superposition over an arbitrary interval}\label{sec:interval-prep}

\subsection{Problem setting and notation}\label{sec:setting}

We represent an integer interval by the half-open notation $[a,a+M)$, whose elements are $a,a+1,\ldots,a+M-1$. We use little-endian indexing. An $n$-qubit computational basis state is written as
\[
  \ket{x}=\ket{q_{n-1}\cdots q_1q_0},
  \qquad x=\sum_{t=0}^{n-1}2^tq_t,
  \quad q_t\in\{0,1\}.
\]
For a nonnegative integer $z=\sum_{t\geq0}z_t2^t$, with $z_t\in\{0,1\}$, we write
\begin{equation}
\label{eq:hamming_wt}
     \wt(z)=\sum_{t\geq0}z_t
\end{equation}
for its binary Hamming weight (the number of nonzero digits in its binary expansion). Throughout, $\land$, $\lor$, and $\oplus$ denote bitwise AND, OR, and XOR, respectively, acting on the binary representations of nonnegative integers. When a mask is applied to a $k$-qubit register, only its lowest $k$ bits are used. A bit mask is an integer whose set-bit positions specify the qubits on which a bitwise operation acts. We say that $m$ is a bitwise submask of $S$ if $m\land S=m$, equivalently if every set-bit position of $m$ is also a set-bit position of $S$.
\Needspace{7\baselineskip}
\begin{samepage}
For a $k$-bit mask $s=\sum_{t=0}^{k-1}2^ts_t$, let $I_t$ and $X_t$ denote
the identity and Pauli-$X$ operators on qubit $q_t$, respectively, and set
$X_t^0=I_t$ and $X_t^1=X_t$. We then define
$X^s:=X_{k-1}^{s_{k-1}}\otimes\cdots\otimes X_0^{s_0}$.
% $ X^s:=\prod_{t=0}^{k-1} X_t^{s_t}$.
In particular, $X^{2^k-1}=X^{\otimes k}$ complements all $k$ qubits.
\end{samepage}
For $1\leq m\leq 2^k$, define
\begin{equation}\label{eq:prefix-state}
  \ket{\psi_m^{(k)}}=\frac{1}{\sqrt m}\sum_{x=0}^{m-1}\ket{x}.
\end{equation}
We call $\ket{\psi_m^{(k)}}$ the \emph{prefix state of length $m$} on $k$ qubits; it is the uniform superposition state on the first $m$ computational-basis labels, $0,\ldots,m-1$.
Unless stated otherwise, the data register is initialized
in the computational-basis state $\ket{0}^{\otimes n}$.

More generally, for a $k$-qubit register and a $k$-bit mask $F=\sum_{j=0}^{k-1}f_j2^j$, where $f_j\in\{0,1\}$,
the state $X^F\ket{\psi_L^{(k)}}$ is the bit-mask transform of the prefix state; its support is obtained by replacing each index $x$ in $[0,L)$ by $x\oplus F$. In particular, when $F=2^k-1$, this transformation gives the suffix state.

A \emph{dyadic block} of size $2^r$ is an aligned interval $[u2^r,(u+1)2^r)$, where $u,r$ are nonnegative integers. Thus
\[
  X^{\otimes k}\ket{\psi_m^{(k)}}
  =\frac{1}{\sqrt m}\sum_{x=0}^{m-1}\ket{2^k-1-x},
\]
which is the uniform quantum superposition state on the interval $[2^k-m,2^k)$.
The Pauli-$Y$ rotation is
\[
  R_y(\theta)=
  \begin{pmatrix}
    \cos(\theta/2)&-\sin(\theta/2)\\
    \sin(\theta/2)&\cos(\theta/2)
  \end{pmatrix}.
\]
For a qubit $c$ and a one-qubit unitary $V$ acting on a distinct target $t$, the notation $\C{q_c=\gamma}{V_t}$ means that $V$ is applied to $t$ exactly when the control qubit has value $\gamma\in\{0,1\}$. We allow positive and negative controls. In the logical gate model used below, a negative control is counted as one singly controlled gate; if only positive controls are available, it can be implemented by conjugating the control wire with Pauli-$X$ gates.

Each one-qubit gate and each singly controlled one-qubit gate counts as one logical gate. We assume that arbitrary real rotation angles are allowed. A singly controlled one-qubit unitary admits an exact decomposition into a constant number of one-qubit gates and CNOT gates. Thus the logical count and the count in a continuous-parameter one- and two-qubit gate model differ by at most a constant factor. The upper bound $5n-9$ refers to the logical model defined here; it is not a CNOT count. Synthesis over a fixed discrete universal gate set may require approximation of arbitrary rotation angles; that synthesis overhead and its approximation error are not included in our count.

\subsection{Uniform prefix-state preparation}
\label{sec:uniform-prefix-preparation}

We use the uniform-superposition preparation algorithm of
Ref.~\cite{shukla2024efficient} (Algorithm~1 therein) to prepare the
prefix state $\ket{\psi_m^{(k)}}$ defined in
Eq.~\eqref{eq:prefix-state}. Its correctness proof and complexity
analysis are given in Sections~2.3 and~2.5 of that paper; we do not
repeat them here. For completeness, we state the algorithm in our
little-endian notation, including the elementary power-of-two cases.

Let $s=\operatorname{wt}(m)$ and let $\ell_0<\ell_1<\cdots<\ell_{s-1}$ be the set-bit positions of $m$.
When $m$ is a power of two, only the Hadamards on the lower
$\log_2m$ qubits are needed. Otherwise $s\geq2$ and the algorithm is as
follows.

\begin{algorithm}[H]
\caption{Uniform prefix-state preparation (restatement of Algorithm~1 of Ref.~\cite{shukla2024efficient})}
\label{alg:prefix-state}
\begin{algorithmic}[1]
\Require $1\leq m\leq2^k$
\Ensure $\ket{\psi_m^{(k)}}$ on $q_{k-1}\cdots q_0$
\If{$m$ is a power of two, say $m=2^v$}
  \State Apply $H$ to $q_0,\ldots,q_{v-1}$; \Return
\EndIf
\State Compute the set-bit positions $\ell_0<\ell_1<\cdots<\ell_{s-1}$ of $m$
\State Apply $X$ to $q_{\ell_1},\ldots,q_{\ell_{s-1}}$
\State Apply $H$ to $q_0,\ldots,q_{\ell_0-1}$
\State $\theta_0\gets-2\arccos\sqrt{2^{\ell_0}/m}$
\State Apply $R_y(\theta_0)$ to $q_{\ell_1}$
\For{$t=\ell_0,\ldots,\ell_1-1$}
  \State Apply $H$ to $q_t$ controlled on $q_{\ell_1}=0$
\EndFor
\State $A\gets2^{\ell_0}$
\For{$j=1,\ldots,s-2$}
  \State $\theta_j\gets-2\arccos\sqrt{2^{\ell_j}/(m-A)}$
  \State Apply $R_y(\theta_j)$ to $q_{\ell_{j+1}}$ controlled on $q_{\ell_j}=0$
  \For{$t=\ell_j,\ldots,\ell_{j+1}-1$}
    \State Apply $H$ to $q_t$ controlled on $q_{\ell_{j+1}}=0$
  \EndFor
  \State $A\gets A+2^{\ell_j}$
\EndFor
\end{algorithmic}
\end{algorithm}

The qubits above the highest set bit of $m$ remain in $\ket0$ when the
register is wider than necessary. This is the only padding convention
needed below. In particular, an aligned prefix interval is prepared by
Algorithm~\ref{alg:prefix-state}, followed by setting its common
high-order bits.

\subsection{A common dyadic-block partition}

Our approach to preparing a uniform state over an arbitrary interval
is to divide the interval into two branches and process them using
a common shared sequence of dyadic blocks. The circuit construction,
presented in the following subsections, uses the same block-selection
positions and expansion gates for both branches, while allowing
their rotation angles and final bit masks to differ. 

Using the aligned dyadic-block convention introduced in
\Cref{sec:setting}, we now construct a common
partition adapted to the two branches.
Consider the target interval
  $I=[a,a+M)$,
and let $b=a+M-1$ denote its last integer. For $M>1$, define
 $ k=\lfloor\log_2(a\mathbin{\oplus}b)\rfloor$.
The binary representations of $a$ and $b$ differ at position
$k$ and agree at all higher positions. Thus, both endpoints
belong to the same aligned dyadic block of size $2^{k+1}$.
Write
\[
  Q=\left\lfloor\frac{a}{2^{k+1}}\right\rfloor,
  \qquad C=2^{k+1}Q,
  \qquad D=C+2^k,
\]
where $C$ is the left endpoint of this parent block and $D$
is its midpoint. 
Throughout this work, ``midpoint'' refers to the midpoint of the
enclosing dyadic parent block, which need not coincide with the
midpoint of the target interval itself.
Since $a\leq b$, the bit of $a$ at position
$k$ is zero and that of $b$ is one.
Therefore,
  $a<D\leq b$,
and the target interval splits into two nonempty branches:
\[
  [a,a+M)=[a,D)\,\cup\,[D,b+1).
\]
We refer to these as the lower and upper branches, respectively,
and denote their lengths by
\begin{equation}\label{eq:branch-lengths}
  L=D-a,\qquad R=b+1-D.
\end{equation}
In particular, $L+R=M$. The qubit $q_k$ distinguishes the
two branches: its value $0$ identifies the lower branch, while
its value $1$ identifies the upper branch.

\begin{example}[Midpoint decomposition with a nonzero prefix]
\label{ex:nonzero-prefix}

Consider the interval $I=[51,59)$ on a six-qubit register.
Here $a=51$, $M=8$, and $b=a+M-1=58$. In six-bit
binary notation,
\[
  a=(110011)_2,\qquad b=(111010)_2.
\]
Their bitwise XOR is $a\mathbin{\oplus}b
  =(110011)_2\mathbin{\oplus}(111010)_2
  =(001001)_2$.
The most significant set bit occurs at position $k=3$.
Thus, the two endpoints belong to the same aligned dyadic
parent block of size $2^{k+1}=16$.
The index of this parent block is
$Q=\left\lfloor\frac{51}{16}\right\rfloor=3=(11)_2$,
so its left endpoint and midpoint are
\[
  C=2^4Q=48=(110000)_2,\qquad
  D=C+2^3=56=(111000)_2.
\]
The parent block is $[48,64)$, and the target interval
splits at $D=56$:
 $ [51,59)=[51,56)\,\cup\,[56,59)$.
The branch lengths are
 $ L=D-a=56-51=5$ and
  $R=b+1-D=59-56=3$.
The example  also illustrates that the midpoint used in the
construction is the midpoint of the enclosing dyadic parent
block, not necessarily the midpoint of the target interval.
\end{example}

\vskip 0.5\baselineskip

Next we consider the generic case
  $1\leq L,R<2^k$.
The aligned-prefix, aligned-suffix, and full dyadic-block cases
are treated separately in \Cref{subsec:special-cases}.
The two branch lengths generally have different binary
decompositions. Constructing their circuits independently would
therefore require different block-selection positions and
potentially different expansion gates. Instead, we seek a
single collection of dyadic blocks that accommodates both
branches. To obtain it, we take the bitwise OR of their lengths:
\begin{equation}\label{eq:partition-S}
  S=L\mathbin{\lor}R.
\end{equation}
Every set-bit position of either $L$ or $R$ is then also a
set-bit position of $S$. Thus, the binary expansion of $S$
identifies all dyadic block sizes that may be needed by either
branch, allowing both branches to use the same block-selection
positions.
Write the binary expansion of $S$ as
\begin{equation}\label{eq:partition-binary}
  S=\sum_{j=0}^{h-1}2^{\lambda_j},
  \qquad
  \lambda_0>\lambda_1>\cdots>\lambda_{h-1}=v,
\end{equation}
where $h=\wt(S)$ is the Hamming weight of $S$. Define
\begin{equation}\label{eq:partition-blocks}
  A_j=\sum_{i=0}^{j-1}2^{\lambda_i},
  \qquad
  B_j=[A_j,A_j+2^{\lambda_j})
  \quad (0\leq j<h).
\end{equation}
Each $B_j$ is an aligned dyadic block of size $2^{\lambda_j}$.
Indeed, every term in $A_j$ is divisible by $2^{\lambda_j}$,
so its left endpoint is a multiple of the block size.
Moreover, the blocks $B_0,\ldots,B_{h-1}$ are pairwise disjoint
and partition $[0,S)$.

Since $L$ and $R$ are bitwise submasks of $S$, every set-bit
position of either branch is among the block-selection positions
$\lambda_j$. Thus, the two branches can be represented using
the same dyadic blocks, with each branch selecting a different
subcollection of them. We call the construction that processes
these blocks in decreasing order of size the \emph{direct
construction}. The reflected construction, defined in
\Cref{sec:reflected-construction}, processes the
reflected blocks in increasing order of size. These names
distinguish two circuit implementations of the same target state.

For each branch length $m\in\{L,R\}$, define the residual lengths
and block indicators recursively by
\begin{equation}\label{eq:branch-residual}
 r_0(m)=m,\qquad
 b_j(m)=\left\lfloor\frac{r_j(m)}{2^{\lambda_j}}\right\rfloor,
 \qquad
 r_{j+1}(m)=r_j(m)-b_j(m)2^{\lambda_j},
 \quad 0\leq j<h.
\end{equation}
Since the set-bit positions of $m$ are among
$\lambda_0,\ldots,\lambda_{h-1}$, every $b_j(m)$ belongs to
$\{0,1\}$. Furthermore, for $0\leq j\leq h$,
\[
 r_j(m)=\sum_{t=j}^{h-1}b_t(m)2^{\lambda_t},
\]
where the sum is empty when $j=h$. In particular,
$r_h(m)=0$. Thus, $r_j(m)$ is the part of branch length $m$
that remains before stage $j$.

The indicator $b_j(m)$ records whether the block $B_j$ is
selected by branch $m$. If $b_j(m)=1$, the block is included
and its size is subtracted from the residual. If $b_j(m)=0$,
the block is skipped and the entire residual is carried forward.
When $r_j(m)=0$, the branch has already terminated and no
further operation is required. At the final position
$\lambda_{h-1}=v$, an active residual is exactly $2^v$.
Hence $b_{h-1}(m)$ simply records whether the last block is
present. No rotation is needed at this final position: the
Hadamards on $q_0,\ldots,q_{v-1}$ already prepare its
low-order superposition.

For an active branch and $0\leq j\leq h-2$, choose the rotation
angle
\begin{equation}\label{eq:branch-theta}
 \theta_j(m)=
 \begin{cases}
  \pi, & b_j(m)=0,\\[1mm]
  2\arctan\sqrt{r_{j+1}(m)/2^{\lambda_j}},
       & b_j(m)=1,\ r_{j+1}(m)>0,\\[1mm]
  0, & b_j(m)=1,\ r_{j+1}(m)=0.
 \end{cases}
\end{equation}
These three cases have a direct interpretation. If
$b_j(m)=0$, the current block is absent, and the angle $\pi$
sends the entire active amplitude to the continuation sector
$q_{\lambda_j}=1$. If $b_j(m)=1$ and a remainder is present,
the rotation divides the amplitude between the selected block
and the continuation. Finally, if $b_j(m)=1$ and no remainder
remains, the branch terminates in the current block, so no
rotation is necessary.

The value $1$ used at a skipped position is a temporary
continuation marker. It is retained through the block-expansion
stage and corrected afterward by the XOR masks. Equivalently,
when $b_j(m)=1$, the rotation angle can be written as
\begin{equation}
    \theta_j(m)=
2\arccos\sqrt{\frac{2^{\lambda_j}}{r_j(m)}}.
\end{equation}

To illustrate the construction, consider the interval \(I=[3,11)\). The endpoint split gives branch lengths \(L=5\) and \(R=3\), and the corresponding common dyadic-block partition, in local coordinates, consists of the blocks \([0,4)\), \([4,6)\), and \([6,7)\). The lower branch selects the first and third blocks, whereas the upper branch selects the second and third. The residual lengths, block-selection indicators, rotation angles, and resulting circuit are worked out in detail in \Cref{subsec:worked-example}.

\begin{lemma}[Preparation over the common dyadic-block partition]
\label{lem:common-partition}
For each $m\in\{L,R\}$, 
use the shared gate locations determined
by $\lambda_0,\ldots,\lambda_{h-1}$ in
Eq.~\eqref{eq:partition-binary}, with the branch-dependent
rotation angles $\theta_j(m)$ from Eq.~\eqref{eq:branch-theta}.
The ordered gate construction is as follows:
\begin{enumerate}
\item Apply Hadamards to $q_0,\ldots,q_{v-1}$.
\item If $h\geq2$, apply a rotation with angle $\theta_0(m)$
      to $q_{\lambda_0}$.
\item For $1\leq j\leq h-2$, apply the rotation with angle
      $\theta_j(m)$ to $q_{\lambda_j}$, controlled on
      $q_{\lambda_{j-1}}=1$.
\item Apply controlled-Hadamards in the order
      $j=h-2,h-3,\ldots,0$. At stage $j$, apply $H$ to
      each $q_t$, for
      $t=\lambda_{j+1},\ldots,\lambda_j-1$,
      controlled on $q_{\lambda_j}=0$.
\end{enumerate}
Rotations with zero angle are omitted. When $h=1$, the
initial Hadamards alone prepare the state.
The resulting lower-register state is
\begin{equation}\label{eq:phi-m}
  \ket{\Phi_m^S}
  =\frac1{\sqrt m}
  \sum_{\substack{0\leq j<h\\b_j(m)=1}}
  \sum_{x\in B_j}\ket{x}.
\end{equation}
In particular, the state is uniform over the union of the
blocks selected by branch $m$.

\end{lemma}

\subsection{Interval-preparation circuit based on a common dyadic-block partition}

The preceding subsection introduced a common dyadic-block partition
for two branch lengths. We now connect this construction to the
original interval and explain how it leads to the desired quantum
circuit. The main idea is to split the interval into a lower and an
upper branch, prepare both branches using the same block-selection
structure, and then apply branch-dependent bitwise XOR operations
to obtain the required support.

The singleton case $M=1$ is prepared directly by applying Pauli-$X$
gates at the set-bit positions of $a$. We therefore assume $M>1$
throughout this subsection. Let
 $ I=[a,a+M)$, $b=a+M-1$,
where $b$ is the last integer in the interval. We first recall the
midpoint decomposition used to identify the two branches. 
Recall that 
$k=\lfloor\log_2(a\mathbin{\oplus}b)\rfloor$.
We also recall the notation:
\begin{equation}\label{eq:midpoint-common}
Q=\left\lfloor\frac{a}{2^{k+1}}\right\rfloor,
 C=2^{k+1}Q,
 D=C+2^k,
 L=D-a,
 R=b+1-D.
\end{equation}
The target interval is the disjoint union
$[a,a+M)=[a,D)\,\cup\,[D,b+1)$
with $L+R=M$. Equivalently, the two branches are
$[a,D)=[C+2^k-L,C+2^k)$ and
$ [D,b+1)=[C+2^k,C+2^k+R)$.
The common high-order bits encode $Q$. We call $q_k$ the
\emph{branch-selector qubit}: its value $0$ identifies the lower
branch $[a,D)$, while its value $1$ identifies the upper branch
$[D,b+1)$.

This decomposition gives a useful state-level description of the
target. Put $F=2^k-1$, and let $\ket{\psi_m^{(k)}}$ denote the
uniform prefix state defined in Eq.~\eqref{eq:prefix-state}.
For a $k$-bit register, $X^F$ denotes the bitwise complement,
obtained by applying $X$ to all its qubits. It maps the prefix
support $\{0,\ldots,L-1\}$ to the uppermost $L$ values
$\{2^k-L,\ldots,2^k-1\}$. The target state can therefore be
written as
\begin{equation}\label{eq:midpoint-state-decomposition}
\ket{\psi_{a,M}^{(n)}}=
\ket{Q}_{\mathrm{high}}\otimes\left(
 \sqrt{\frac LM}\ket0_{q_k}X^F\ket{\psi_L^{(k)}}
 +\sqrt{\frac RM}\ket1_{q_k}\ket{\psi_R^{(k)}}
\right).
\end{equation}

The two branches have probabilities $L/M$ and $R/M$,
respectively, and hence the correct relative weights. Since
$L+R=M$, the resulting state is normalized. Thus, the
decomposition has the desired support and amplitudes: the
lower branch is a prefix state transformed by the bit mask
$F$, whereas the upper branch is an ordinary prefix state.

Equation~\eqref{eq:midpoint-state-decomposition} also suggests a
straightforward approach: prepare the two prefix states separately,
combine them using the selector qubit, and reflect the lower branch.
However, independent prefix-state preparations generally use
different block-selection positions and different expansion
patterns. Implementing two separate circuits would therefore fail
to exploit the common binary structure of the branch lengths.
Instead, we use the common dyadic-block partition introduced in
the preceding subsection. The bitwise OR
$S=L\mathbin{\lor}R$
identifies a common set of block-selection positions for both
branches. We first prepare the corresponding common-partition state
$\ket{\Phi_m^S}$ for each branch length $m\in\{L,R\}$, using the
same gate positions and the appropriate branch-dependent rotation
angles. The resulting block structure is then converted to the
required prefix coordinates by the XOR masks of
Lemma~\ref{lem:xor-conversion}.

More precisely, let $\sigma_L$ and $\sigma_R$ be the masks defined
in Eq.~\eqref{eq:sigma-mask}. They convert the common-partition
coordinates into the ordinary prefix coordinates:
\begin{equation}
    X^{\sigma_L}\ket{\Phi_L^S}=\ket{\psi_L^{(k)}},
 \qquad
 X^{\sigma_R}\ket{\Phi_R^S}=\ket{\psi_R^{(k)}}.
\end{equation}
To obtain the desired lower branch, we must additionally reflect
the lower prefix state. Accordingly, the two final masks are
\begin{equation} \label{eq:final-branch-masks}
\tau_L=F\mathbin{\oplus}\sigma_L,
 \qquad
 \tau_R=\sigma_R. 
\end{equation}
For $m\in\{L,R\}$, let $v_m=\nu_2(m)$, where $\nu_2(m)=\max\{v\geq0:2^v\mid m\}$ be the $2$-adic valuation of $m$.
The lowest $v_m$ qubits of $\ket{\Phi_m^S}$ are in
the state $\ket{+}^{\otimes v_m}$. Therefore, applying
Pauli-$X$ to any of these qubits leaves the branch state
unchanged. We may therefore omit the lowest $v_m$ bits
of the final branch mask. Define the reduced branch masks
by
\begin{equation}\label{eq:reduced-branch-masks}
  \widehat{\tau}_L
    =\tau_L\mathbin{\land}(2^k-2^{v_L}),
  \qquad
  \widehat{\tau}_R
    =\tau_R\mathbin{\land}(2^k-2^{v_R}).
\end{equation}
In particular,
\[
  X^{\widehat{\tau}_m}\ket{\Phi_m^S}
  =
  X^{\tau_m}\ket{\Phi_m^S},
  \qquad m\in\{L,R\}.
\]
Since bitwise XOR masks compose by XOR, these give
\begin{equation}
    X^{\tau_L}\ket{\Phi_L^S}
 =X^F\ket{\psi_L^{(k)}},
 \qquad
 X^{\tau_R}\ket{\Phi_R^S}
 =\ket{\psi_R^{(k)}}.
\end{equation}
Thus, the common partition handles the preparation of both
branches, while the final masks account for their different
coordinate locations and the reflection of the lower branch.
In general, these coordinate corrections cannot be omitted.

\subsection{Paired-angle implementation of branch-dependent rotations}
\label{subsec:paired-angle-rotations}

The common partition uses the same block-selection qubits in both selector branches, but the rotation angles can differ. At stage $j=0$, the selector $q_k$ is the only control
that distinguishes the two angles. The two branch rotations
are therefore implemented using at most two operations:
one unconditional baseline rotation and, when the angles
are nonzero and unequal, one selector-controlled correction.
If the angles agree, only the baseline rotation is needed.
If one angle is zero, a single selector-controlled rotation
suffices.

At every stage after the first, $j\geq1$, the target qubit $t=q_{\lambda_j}$ has not yet been acted upon. The initial Hadamards act only on positions below $v$, all earlier rotations have distinct targets, and the block-expansion stage has not yet begun. Hence $t$ is still in $\ket0$ when its stage-$j$ rotation is applied. This invariant permits a three-gate implementation that simultaneously accounts for both branch angles. In particular, this construction avoids the need to synthesize a general doubly controlled rotation.

\begin{lemma}[Three-gate paired-angle identity on a zero target]\label{lem:paired-angle}
Let $c,d,t$ be distinct qubits, and suppose the target is in $\ket0_t$. Let $\theta_0,\theta_1\in\mathbb R$ and put
\[
  \Delta=\theta_1-\theta_0,
  \qquad \overline\theta=\frac{\theta_0+\theta_1}{2}.
\]
Apply the following three singly controlled gates in the order listed:
\begin{equation}
    \label{eq:paired-angle-first}
   C_{d=1}\!\left(R_y(-\Delta/2)_t\right), \qquad
   C_{c=1}(Z_t),
  \qquad  C_{d=1}\!\left(R_y(\overline\theta)_t\right).
\end{equation} 
The resulting action on the target is
\[
\begin{array}{c|c|c}
 c&d&\text{output target state}\\ \hline
 0&0&\ket0\\
 1&0&\ket0\\
 0&1&R_y(\theta_0)\ket0\\
 1&1&R_y(\theta_1)\ket0.
\end{array}
\]
The controls are unchanged. The identity is exact on the stated input subspace, including when the controls are in coherent superposition or entangled with other systems.
\end{lemma}

At stage $j\geq1$, take $c=q_k$, $d=q_{\lambda_{j-1}}$, $t=q_{\lambda_j}$, $\theta_0=\theta_j(L)$, and $\theta_1=\theta_j(R)$. If both branches are active and their angles differ, Lemma~\ref{lem:paired-angle} implements the two required branch rotations with three singly controlled gates in total. This includes the case in which one of the two angles is zero. If the angles agree, one controlled rotation on $d=1$ is enough. If only one branch remains active, its angle is implemented by one controlled rotation on $d=1$; the terminated branch has $d=0$ and is unaffected. If both branches have terminated, the stage is omitted.

The three-gate identity is specific to state preparation with a target initially in $\ket0$; it is not a general two-control unitary decomposition. In the inactive sector $c=1,d=0$, the middle $Z$ is harmless because the target is known to be $\ket0$. This condition holds at every subsequent block-selection stage of the common dyadic-block construction.

The paired-angle construction can also be viewed as a
common-angle implementation with a branch-dependent sign
correction. A direct common-angle approach would apply
$R_y(\theta_0)$ controlled on $d=1$, followed by a correction
$R_y(\theta_1-\theta_0)$ controlled jointly on $c=d=1$.
The latter is a doubly controlled rotation. The identity in
Lemma~\ref{lem:paired-angle} avoids this operation by using
the relation
\[
  ZR_y(\alpha)\ket0=R_y(-\alpha)\ket0.
\]
It replaces the joint-control correction by a controlled
phase operation, which is harmless in the inactive sector
because the target is initially $\ket0$. Thus, arbitrary
branch-dependent angles are implemented using only three
singly controlled gates.

\subsection{XOR conversion from blocks of the common partition to prefix blocks}

For a positive integer $m$, let $v_m=\nu_2(m)=\max\{v\geq0:2^v\mid m\}$ be its $2$-adic valuation. Define the mask
\begin{equation}\label{eq:sigma-mask}
  \sigma_m=(S\mathbin{\oplus}m)\mathbin{\land}\bigl(2^k-2^{v_m+1}\bigr),
\end{equation}
where the bitwise operations are restricted to the lower $k$ bits. Thus $\sigma_m$ toggles the set-bit positions of $S$ that are absent from $m$ only at positions strictly above $v_m$.

\begin{lemma}[XOR conversion of selected blocks of the common partition]\label{lem:xor-conversion}
For $m\in\{L,R\}$,
\begin{equation}\label{eq:xor-prefix}
  X^{\sigma_m}\ket{\Phi_m^S}=\ket{\psi_m^{(k)}}.
\end{equation}
Moreover, $\sigma_L\mathbin{\land}\sigma_R=0$.
\end{lemma}

\subsection{Aligned prefixes, aligned suffixes, and dyadic intervals}
\label{subsec:special-cases}

The endpoint split identifies two cases that can be handled directly
using the uniform-superposition preparation algorithm reviewed in
\Cref{sec:uniform-prefix-preparation}. Thus, before considering the general
common-partition construction, we isolate these cases.

For $M>1$, put $r_M=\lceil\log_2 M\rceil$, where $2^{r_M-1}<M\leq 2^{r_M}$.
The quantity $2^{r_M}$ is the smallest dyadic block width that can
contain an interval of length $M$. In terms of the endpoint split
defined in \eqref{eq:midpoint-common}, the two alignment conditions
are
\begin{align}
    a\equiv 0\pmod{2^{r_M}}
    &\quad\Longleftrightarrow\quad L=2^k,
    \label{eq:aligned-prefix-equivalence}\\
    a+M\equiv 0\pmod{2^{r_M}}
    &\quad\Longleftrightarrow\quad R=2^k.
    \label{eq:aligned-suffix-equivalence}
\end{align}
In the first case the interval begins at the left endpoint of its
aligned dyadic parent block; in the second it ends at the right
endpoint.

\paragraph{Aligned prefix.}
If $L=2^k$, the interval begins at the left endpoint of the parent
block. Relative to its common high-order prefix $Q$, the desired
state on the lower $k+1$ qubits is precisely the prefix state
\[
    \ket{\psi_M^{(k+1)}}=
    \frac{1}{\sqrt M}\sum_{x=0}^{M-1}\ket{x}.
\]
It can therefore be prepared directly by
Algorithm~\ref{alg:prefix-state} on $q_k,\ldots,q_0$. The common
high-order bits are then set to $Q$. We use the earlier
uniform-superposition preparation algorithm without modification.

\paragraph{Aligned suffix.}
If $R=2^k$, the interval ends at the right endpoint of the parent
block. Its relative support on the lower $k+1$ qubits is
   $ [2^{k+1}-M,\,2^{k+1})$.
This case is also a direct consequence of the prefix-state
preparation algorithm.
Indeed, bitwise complementation of a $(k+1)$-bit register maps
$x$ to $2^{k+1}-1-x$, and hence
\begin{equation}
X^{\otimes(k+1)}\ket{\psi_M^{(k+1)}}
    =
    \frac{1}{\sqrt M}
    \sum_{x=2^{k+1}-M}^{2^{k+1}-1}\ket{x}.
    \label{eq:suffix-as-complement}
\end{equation}
Therefore, one may prepare the prefix state using
Algorithm~\ref{alg:prefix-state} and then complement all $k+1$ qubits.

For the gate-count comparison below, it is useful to obtain a
complement-adapted descending-block realization of the prefix
state. This realization uses the block-selection and
block-expansion construction developed in
Lemma~\ref{lem:common-partition}, specialized to a single
length $M$ with common partition parameter $S=M$.
Write
\[
    M=2^{v_M}N_M,\qquad
    N_M=\sum_{j=0}^{s_M-1}2^{d_j},\qquad
    d_0>d_1>\cdots>d_{s_M-1}=0,
\]
where $N_M$ is odd and $s_M=\operatorname{wt}(N_M)$. The
corresponding dyadic block sizes are
$2^{v_M+d_j}$. Since every set bit of $M$ is present in
the partition, every block is selected, and no
branch-dependent XOR correction is required. The resulting
descending-block circuit prepares the prefix state
$\ket{\psi_M^{(k+1)}}$ on $q_k,\ldots,q_0$.

We obtain the aligned-suffix circuit by complementing this
output. Instead of applying all $k+1$ bit flips after state
preparation, we absorb the complement on the highest
block-selection qubit into the first rotation and reverse
the polarity of the subsequent controls involving that
qubit. The remaining necessary output flips are applied
explicitly. This gives the following complement-adapted
realization.

Let $\lambda_j=v_M+d_j$ and, for
$0\leq j\leq s_M-2$, define
\[
    \rho_j^{(M)}
    =
    M-\sum_{i=0}^{j-1}2^{v_M+d_i},
    \qquad
    \theta_j
    =
    2\arccos\sqrt{
      \frac{2^{v_M+d_j}}{\rho_j^{(M)}}
    }.
\]
These are the block-selection angles of the descending-block
construction specialized to the single length $M$.

\begin{lemma}[Complement-adapted descending-block circuit for an aligned suffix]
\label{lem:aligned-suffix}
Suppose $2^k<M<2^{k+1}$. The state
\[
    \frac{1}{\sqrt M}
    \sum_{x=2^{k+1}-M}^{2^{k+1}-1}\ket{x}
\]
can be prepared on $q_k,\ldots,q_0$ by the following modification
of the prefix-state preparation circuit:
\begin{enumerate}[label=(\roman*),leftmargin=2.2em]
    \item Apply $H$ to $q_0,\ldots,q_{v_M-1}$.
    \item Apply $R_y(\pi-\theta_0)$ to $q_k$.
    \item For $j=1,\ldots,s_M-2$, apply $R_y(\theta_j)$ to
    $q_{v_M+d_j}$, controlled on $q_k=0$ when $j=1$, and on
    $q_{v_M+d_{j-1}}=1$ when $j\geq2$.
    \item For $j=s_M-2,s_M-3,\ldots,1$ and
    $t=d_{j+1},\ldots,d_j-1$, apply $H$ to $q_{v_M+t}$,
    controlled on $q_{v_M+d_j}=0$. In addition, for
    $t=d_1,\ldots,d_0-1$, apply $H$ to $q_{v_M+t}$,
    controlled on $q_k=1$.
    \item Apply $X$ to $q_{v_M},\ldots,q_{k-1}$.
\end{enumerate}
The common high-order bits above $q_k$ are set to $Q$.
\end{lemma}

The direct complement-adapted suffix circuit uses $v_M$
Hadamards, $s_M-1$ rotations, $k-v_M$ controlled Hadamards, and
$k-v_M$ final bit flips. Its logical gate count on the lower
$K=k+1$ qubits is therefore
\begin{equation}
    C_1
    =v_M+(s_M-1)+2(k-v_M)
    =K+h_M+s_M-3,
    \qquad h_M=k-v_M+1.
    \label{eq:direct-suffix-C1}
\end{equation}
The reflected suffix construction and the comparison of the two
suffix circuits are given in
\Cref{subsec:reflected-suffix}.

If $M=2^{k+1}$, the interval is the entire parent block and is
prepared simply by applying $H$ to all $k+1$ lower qubits. This
case is already covered by the aligned-prefix branch.
Dyadic intervals are consequently a particularly simple instance
of the aligned-prefix case: if $M=2^r$ and $a$ is divisible by
$2^r$, the lower $r$ qubits are prepared in
$\ket{+}^{\otimes r}$, while the higher qubits encode the aligned
block index. The common-partition construction is therefore
needed only for intervals that are neither aligned prefixes nor
aligned suffixes.

\begin{algorithm}[H]
\caption{Interval preparation using a common dyadic-block partition}
\label{alg:interval-main}
\begin{algorithmic}[1]\small
\Require $n\geq1$ and integers $0\leq a<a+M\leq2^n$
\Ensure $\ket{\psi_{a,M}^{(n)}}$ on $q_{n-1}\cdots q_0$
\If{$M=1$}
  \State Apply $X$ to each $q_t$ whose bit in $a$ is one; \Return
\EndIf
\State Compute $b=a+M-1$, $k=\lfloor\log_2(a\oplus b)\rfloor$, and $C,D,L,R,Q$ from \eqref{eq:midpoint-common}
\If{$L=2^k$}
  \State Run Algorithm~\ref{alg:prefix-state} on $q_k,\ldots,q_0$; set high bits to $Q$; \Return
\ElsIf{$R=2^k$}
  \State Apply the direct complement-adapted suffix circuit of Lemma~\ref{lem:aligned-suffix} on $q_k,\ldots,q_0$
  \State Set high bits to $Q$; \Return
\EndIf
\State $S\gets L\mathbin{\lor}R$; write $S=\sum_{j=0}^{h-1}2^{\lambda_j}$ with $\lambda_0>\cdots>\lambda_{h-1}=v$
\State Apply $R_y(2\arccos\sqrt{L/M})$ to selector $q_k$
\State Apply $H$ to $q_0,\ldots,q_{v-1}$
\For{$j=0,\ldots,h-2$}
  \State Determine activity from $r_j(L),r_j(R)>0$; compute $\theta_j(m)$ only for active branches
  \If{$j=0$}
    \If{$\theta_0(L)=\theta_0(R)$}
      \State Apply $R_y(\theta_0(L))$ to $q_{\lambda_0}$
    \ElsIf{$\theta_0(L)=0$}
      \State Apply $R_y(\theta_0(R))$ to $q_{\lambda_0}$ controlled on $q_k=1$
    \ElsIf{$\theta_0(R)=0$}
      \State Apply $R_y(\theta_0(L))$ to $q_{\lambda_0}$ controlled on $q_k=0$
    \Else
      \State Apply $R_y(\theta_0(L))$ to $q_{\lambda_0}$
      \State Apply $R_y(\theta_0(R)-\theta_0(L))$ to $q_{\lambda_0}$ controlled on $q_k=1$
    \EndIf
  \ElsIf{both branches active and their angles differ}
    \State $d\gets q_{\lambda_{j-1}}$, $t\gets q_{\lambda_j}$, $\Delta_j\gets\theta_j(R)-\theta_j(L)$
    \State Apply $R_y(-\Delta_j/2)$ to $t$ controlled on $d=1$
    \State Apply $Z$ to $t$ controlled on $q_k=1$
    \State Apply $R_y((\theta_j(L)+\theta_j(R))/2)$ to $t$ controlled on $d=1$
  \ElsIf{at least one branch remains active}
    \State $\theta\gets\theta_j(L)$ if the lower branch is active, and $\theta\gets\theta_j(R)$ otherwise
    \State Apply $R_y(\theta)$ to $q_{\lambda_j}$ controlled on $q_{\lambda_{j-1}}=1$
  \EndIf
\EndFor
\For{$j=h-2,h-3,\ldots,0$}
  \For{$t=\lambda_{j+1},\ldots,\lambda_j-1$}
    \State Apply $H$ to $q_t$ controlled on $q_{\lambda_j}=0$
  \EndFor
\EndFor
\State Compute $\tau_L=(2^k-1)\oplus\sigma_L$ and $\tau_R=\sigma_R$ from \eqref{eq:sigma-mask}
\State Set $\widehat\tau_L=\tau_L\mathbin{\land}(2^k-2^{v_L})$ and $\widehat\tau_R=\tau_R\mathbin{\land}(2^k-2^{v_R})$, where $v_L=\nu_2(L)$ and $v_R=\nu_2(R)$
\For{$t=0,\ldots,k-1$}
  \If{bit $t$ is set in both $\widehat\tau_L$ and $\widehat\tau_R$}
    \State Apply $X$ to $q_t$
  \ElsIf{bit $t$ is set only in $\widehat\tau_L$}
    \State Apply $X$ to $q_t$ controlled on $q_k=0$
  \ElsIf{bit $t$ is set only in $\widehat\tau_R$}
    \State Apply $X$ to $q_t$ controlled on $q_k=1$
  \EndIf
\EndFor
\State Set the bits above $q_k$ to the common prefix $Q$
\end{algorithmic}
\end{algorithm}

\begin{samepage}
\begin{theorem}[Exact interval preparation using a common dyadic-block partition]\label{thm:interval-main}
Algorithm~\ref{alg:interval-main} prepares
\[
 \ket{\psi_{a,M}^{(n)}}=\frac1{\sqrt M}\sum_{x=a}^{a+M-1}\ket{x}
\]
exactly for every $0\leq a<a+M\leq2^n$. It uses no ancillary qubits and no gates with more than one control.
\end{theorem}

\end{samepage}

\section{Correctness of the direct construction}\label{sec:direct-correctness}

The statements of the auxiliary results used in Algorithm~\ref{alg:interval-main}
were given in \Cref{sec:interval-prep}. We now prove
Lemmas~\ref{lem:common-partition}, 
\ref{lem:paired-angle}, \ref{lem:xor-conversion} and \ref{lem:aligned-suffix}, and then establish
Theorem~\ref{thm:interval-main}. The proofs are presented in the same
order in which these results support the direct construction.

\subsection{Correctness of the common dyadic-block preparation}

\begin{proof}[Proof of Lemma~\ref{lem:common-partition}]
Fix a branch length $m\in\{L,R\}$. We first analyze the
block-selection rotations and then the block-expansion gates.

\medskip
\noindent
\emph{Block selection.}
Initially, the Hadamards on $q_0,\ldots,q_{v-1}$ prepare
a uniform superposition over the $2^v$ low-order basis
states. The remaining block-selection qubits are still
in $\ket0$. At stage $j$, the component corresponding to
the residual length $r_j(m)$ has squared norm
 $ \frac{r_j(m)}{m}$,
and the target qubit $q_{\lambda_j}$ is still in $\ket0$
on this component. For $j\geq1$, this active component is
identified by the preceding continuation marker
$q_{\lambda_{j-1}}=1$.

If $b_j(m)=0$, the current block is absent. The angle
$\theta_j(m)=\pi$ therefore sends the entire active
component to $q_{\lambda_j}=1$, so the residual continues
unchanged.
If $b_j(m)=1$, then
 $ r_j(m)=2^{\lambda_j}+r_{j+1}(m)$.
For $r_{j+1}(m)>0$, the chosen angle gives
\[
 R_y(\theta_j(m))\ket0
 =
 \sqrt{\frac{2^{\lambda_j}}{r_j(m)}}\ket0
 +
 \sqrt{\frac{r_{j+1}(m)}{r_j(m)}}\ket1.
\]
Thus, the component with $q_{\lambda_j}=0$ is assigned
to the block $B_j$, while the component with
$q_{\lambda_j}=1$ continues to the next stage. Their
squared norms are, respectively,
 $\frac{2^{\lambda_j}}{m}$
 and
 $\frac{r_{j+1}(m)}{m}$.
 If $r_{j+1}(m)=0$, the angle is zero and the entire
remaining amplitude is assigned to $B_j$; the branch
terminates.

It follows that each selected block $B_j$ receives total
probability $2^{\lambda_j}/m$, and that the residual
probability after stage $j$ is $r_{j+1}(m)/m$.
In particular, the selected blocks have total size
\[
  \sum_{j=0}^{h-1}b_j(m)2^{\lambda_j}=m.
\]

\medskip
\noindent
\emph{Block expansion.}
It remains to show that the controlled-Hadamard gates
expand each selected block uniformly, without affecting
the other selected blocks.
Suppose $B_i$ is selected. Its block-selection qubits
have the pattern
\[
 q_{\lambda_j}=
 \begin{cases}
  1,&j<i,\\
  0,&j\geq i.
 \end{cases}
\]
Indeed, the value $1$ at an earlier position records that
the corresponding block was skipped and the residual
continued, whereas the value $0$ at position $i$ marks
the selected block. All subsequent block-selection
positions remain zero.

The controlled-Hadamard gates are applied in decreasing
order of $j$, from the smallest blocks towards the largest.
At stage $j$, the gates act precisely when
$q_{\lambda_j}=0$. Therefore, for the block $B_i$,
the expansion gates at stages $j\geq i$ act, whereas
those at stages $j<i$ do not.
The target positions at these active stages are
 $[\lambda_{j+1},\lambda_j),
 \qquad i\leq j\leq h-2$.
Together with the initial Hadamard positions
$[0,v)$, these intervals form the disjoint decomposition
\[
 [0,\lambda_i)
 =
 [0,v)\,\cup
 \bigcup_{j=i}^{h-2}[\lambda_{j+1},\lambda_j).
\]
Thus, exactly $\lambda_i$ Hadamard gates act on the
lower-order qubits of the component associated with $B_i$.
Before expansion, this component has squared norm
$2^{\lambda_i}/m$. Its amplitude is therefore distributed
uniformly over the $2^{\lambda_i}$ basis states of $B_i$,
with amplitude
\[
 \sqrt{\frac{2^{\lambda_i}}{m}}\,
 2^{-\lambda_i/2}
 =\frac1{\sqrt m}
\]
on each state.
The selected blocks are pairwise disjoint, and their
total size is $m$. Hence every computational-basis state
in their union has amplitude $1/\sqrt m$, proving
\eqref{eq:phi-m}.
\end{proof}

\subsection{Correctness of the paired-angle implementation}

\begin{proof}[Proof of Lemma~\ref{lem:paired-angle}]
When $d=0$, the first and third gates are inactive. If $c=0$, the middle gate is also inactive; if $c=1$, it applies $Z$ to $\ket0$, which leaves the target unchanged. When $d=1$ and $c=0$, only the first and third rotations act, giving
\[
 R_y(\overline\theta)R_y(-\Delta/2)\ket0
 =R_y(\overline\theta-\Delta/2)\ket0
 =R_y(\theta_0)\ket0.
\]
When $d=c=1$, use
\[
 ZR_y(-\Delta/2)\ket0=R_y(\Delta/2)\ket0,
\]
so the final rotation gives $R_y(\overline\theta+\Delta/2)\ket0=R_y(\theta_1)\ket0$. The controls are never targets, so the blockwise identities extend by linearity to coherent superpositions and entangled inputs.
\end{proof}

\subsection{Correctness of XOR conversion}

\begin{proof}[Proof of Lemma~\ref{lem:xor-conversion}]
Consider a selected block $B_j$, so that $b_j(m)=1$ and hence $\lambda_j\geq v_m$. Its starting point is $A_j=\sum_{i<j}2^{\lambda_i}$. In the binary representation of $A_j$, the set-bit positions of $A_j$ above $\lambda_j$ are exactly those of $S$ above $\lambda_j$, the bit at $\lambda_j$ is zero in $A_j$, and the lower $\lambda_j$ positions are zero. Above $\lambda_j$, the mask $\sigma_m$ toggles exactly those set-bit positions of $S$ that are absent from $m$. It does not toggle the bit at $\lambda_j$. Therefore the high-order part of the image block is the integer represented by the set bits of $m$ above $\lambda_j$, with the bit at position $\lambda_j$ remaining zero. Mask bits below $\lambda_j$ only permute the $2^{\lambda_j}$ possible low-order strings. It follows that
\[
 B_j\mathbin{\oplus}\sigma_m
 =\left[P_j(m),P_j(m)+2^{\lambda_j}\right),
 \qquad
 P_j(m)=\sum_{i<j}b_i(m)2^{\lambda_i}.
\]
As $j$ ranges over the set bits of $m$, these are exactly the consecutive dyadic blocks in the binary decomposition of $[0,m)$. This proves \eqref{eq:xor-prefix}.

For the disjointness assertion, if a bit of $\sigma_L$ is one, then the corresponding bit of $S$ is one, the bit of $L$ is zero, and therefore the bit of $R$ is one. The corresponding bit of $\sigma_R$ is therefore zero. Thus the masks are disjoint.
\end{proof}

\subsection{Correctness of the aligned-suffix specialization}

\begin{proof}[Proof of Lemma~\ref{lem:aligned-suffix}]
Consider the single-length descending-block construction with
common partition parameter $S=M$. Its binary block positions are
\[
    \lambda_j=v_M+d_j,\qquad 0\leq j<s_M.
\]
Since every set bit of $M$ is included, every dyadic block
is selected. The block-selection and block-expansion
operations therefore prepare
\[
    \ket{\psi_M^{(k+1)}}
    =\frac1{\sqrt M}\sum_{x=0}^{M-1}\ket{x}.
\]
There is no nontrivial XOR correction in this specialization.

We now complement the output. Since
$v_M+d_0=k$, the highest block-selection qubit is $q_k$.
The complement on this qubit can be absorbed into the first
rotation using
\[
    X R_y(\theta_0)\ket0
    =R_y(\pi-\theta_0)\ket0.
\]
This changes the first rotation to the one in part~(ii).
It also reverses the polarity of every subsequent control
involving $q_k$: the continuation control at the first
subsequent selection stage changes from $q_k=1$ to
$q_k=0$, while the expansion control associated with the
largest block changes from $q_k=0$ to $q_k=1$.
The controls involving the remaining block-selection qubits
retain their original polarities. These are precisely the
control conditions in parts~(iii) and~(iv).

It remains to account for the output bit flips. The lowest
$v_M$ qubits are in $\ket+^{\otimes v_M}$ and are unchanged
by complementation. The remaining output flips are applied
to $q_{v_M},\ldots,q_{k-1}$, as specified in part~(v).
This also correctly complements any expansion target already
in $\ket+$, while mapping an unexpanded $\ket0$ to $\ket1$.

Thus the resulting circuit implements
$X^{\otimes(k+1)}\ket{\psi_M^{(k+1)}}$, which is the
desired aligned-suffix state by
\eqref{eq:suffix-as-complement}.
\end{proof}

\subsection{Proof of the main interval-preparation theorem}

\begin{proof}[Proof of Theorem~\ref{thm:interval-main}]
The singleton case is exact by direct basis-state preparation. The aligned-prefix case is exact by Algorithm~\ref{alg:prefix-state}, and the aligned-suffix case by Lemma~\ref{lem:aligned-suffix}. Consider the remaining case $1\leq L,R<2^k$.
The selector rotation prepares
\[
 \sqrt{\frac LM}\ket0_{q_k}+\sqrt{\frac RM}\ket1_{q_k}.
\]
By Lemma~\ref{lem:common-partition} and Lemma~\ref{lem:paired-angle}, the common dyadic-block partition prepares $\ket{\Phi_L^S}$ and $\ket{\Phi_R^S}$ in the corresponding selector branches. The targets in the paired-angle identity are still in $\ket0$ at their respective stages, so its use is valid. Lemma~\ref{lem:xor-conversion} gives
\[
 X^{\sigma_L}\ket{\Phi_L^S}=\ket{\psi_L^{(k)}},\qquad
 X^{\sigma_R}\ket{\Phi_R^S}=\ket{\psi_R^{(k)}}.
\]
The final masks are $\tau_L=(2^k-1)\oplus\sigma_L$ and $\tau_R=\sigma_R$. Removing their lowest $v_m=\nu_2(m)$ bits does not change the corresponding branch state, because those qubits are in $\ket+$. The mask loop therefore implements the required lower-branch complement and upper-branch prefix.

The resulting lower-register state is
\[
 \sqrt{\frac LM}\ket0X^{\otimes k}\ket{\psi_L^{(k)}}
 +\sqrt{\frac RM}\ket1\ket{\psi_R^{(k)}}.
\]
The first branch is uniform on $[2^k-L,2^k)$ and the second on $[0,R)$. After adjoining the common high-order prefix $Q$, these are exactly $[a,D)$ and $[D,b+1)$. Every occupied basis state has amplitude
\[
 \sqrt{\frac LM}\frac1{\sqrt L}
 =\sqrt{\frac RM}\frac1{\sqrt R}
 =\frac1{\sqrt M}.
\]
The supports are disjoint and their union is $[a,a+M)$. All operations have at most one control and the construction uses only the $n$ data qubits.
\end{proof}

\section{Reflected construction}
\label{sec:reflected-construction}

The direct construction processes the common dyadic blocks in
decreasing order of size. An alternative is obtained by reflecting
the active register coordinates and processing the resulting blocks
in increasing order of size. We refer to this alternative as the
\emph{reflected construction}. It uses the same block-selection
principle and paired-angle identity as the direct construction,
but reverses the block order and the associated control polarities.
Like the direct construction, it is exact, ancilla-free, and uses
only one-qubit and singly controlled one-qubit gates.

Recall that $S=\sum_{j=0}^{h-1}2^{\lambda_j}$, where
$\lambda_0>\lambda_1>\cdots>\lambda_{h-1}=v$. Put
$e_r=\lambda_{h-1-r}$ for $0\leq r<h$, so that
$e_0<e_1<\cdots<e_{h-1}=\lambda_0$. For
$0\leq r<h$, define
\begin{equation}\label{eq:reflected-blocks}
\begin{aligned}
A_r =\sum_{s=0}^{r-1}2^{e_s}, \quad
 D_r =\left[
 2^{\lambda_0+1}-S+A_r,\,
 2^{\lambda_0+1}-S+A_r+2^{e_r}
 \right).
\end{aligned}
\end{equation}
The intervals $D_0,\ldots,D_{h-1}$ are pairwise disjoint
and partition $[2^{\lambda_0+1}-S,2^{\lambda_0+1})$.
Moreover, $D_r$ is the image of $B_{h-1-r}$ under the
active-register reflection $x\mapsto F_{\mathrm{act}}-x$,
where $F_{\mathrm{act}}=2^{\lambda_0+1}-1$. Thus, only the
lowest $\lambda_0+1$ qubits are reflected; any higher
padding qubits remain unchanged.

For each branch length $m\in\{L,R\}$, define the residual
length and the size of the block considered at stage $r$ by
$V_r(m)=\sum_{s=r}^{h-1}b_{h-1-s}(m)2^{e_s}$ and
$P_r(m)=b_{h-1-r}(m)2^{e_r}$, respectively. In particular,
$V_0(m)=m$ and $V_{r+1}(m)=V_r(m)-P_r(m)$. For
$0\leq r\leq h-2$ with $V_r(m)>0$, the corresponding
rotation angle is
\begin{equation}\label{eq:reflected-rotation-angle}
 \vartheta_r(m)=2\arccos\sqrt{\frac{P_r(m)}{V_r(m)}}.
\end{equation}
A zero block gives angle $\pi$, a block containing the entire
remaining mass gives angle zero, and a terminated branch requires
no further operation.

The initial bit-flip set is
\begin{equation}\label{eq:reflected-initial-flips}
 \mathcal I_0=
 \{v,v+1,\ldots,\lambda_0\}
 \setminus\{e_1,\ldots,e_{h-1}\}.
\end{equation}
Apply $X$ to the qubits indexed by $\mathcal I_0$ and
Hadamard gates to $q_0,\ldots,q_{v-1}$. Thus,
$q_{e_1},\ldots,q_{e_{h-1}}$ initially remain in $\ket0$,
whereas the other qubits in the range $[v,\lambda_0]$
are initialized in $\ket1$. The higher padding qubits
remain in $\ket0$.

The branch selector is prepared independently by applying
$R_y(2\arccos\sqrt{L/M})$ to $q_k$, producing
$\sqrt{L/M}\ket0_{q_k}+\sqrt{R/M}\ket1_{q_k}$.
When $h=1$, necessarily $L=R=2^v$, and this operation
reduces to a Hadamard gate.

\paragraph{Block selection.}
The block-selection rotations are applied in the order
$r=0,1,\ldots,h-2$. At stage $r$, the target is
$q_{e_{r+1}}$. There is no block-selection control at the
first stage; at every subsequent stage, the rotation is
controlled on $q_{e_r}=1$.

Before its rotation, the target $q_{e_{r+1}}$ is still in
$\ket0$, since all preceding rotations have distinct targets.
Therefore, the angle $\vartheta_r(m)$ assigns the fraction
$P_r(m)/V_r(m)$ of the active squared norm to the block
$D_r$, leaving the fraction $V_{r+1}(m)/V_r(m)$ for
the continuation.

At stage $r=0$, the two branch-dependent angles are implemented
using at most two gates: one common rotation when the angles
agree, one selector-controlled rotation when one angle is zero,
or a baseline rotation followed by a selector-controlled
correction otherwise. For $r\geq1$, when both branches are
active and their angles differ, apply
Lemma~\ref{lem:paired-angle} with
$c=q_k$, $d=q_{e_r}$, $t=q_{e_{r+1}}$,
$\theta_0=\vartheta_r(L)$, and
$\theta_1=\vartheta_r(R)$.
The zero-target condition of that lemma holds by the target
invariant above. If the angles agree, one controlled rotation
suffices. If only one branch remains active, apply its
rotation; if both branches have terminated, omit the stage.

\paragraph{Block expansion.}
Write $XHX$ for the one-qubit unitary satisfying
$XHX\ket1=\ket+$. The reflected expansion follows from
the conjugation identity
\begin{equation}\label{eq:reflected-conjugation}
 (X_c\otimes X_t)\,
 \mathsf C_{c=0}(H_t)\,
 (X_c\otimes X_t)
 =
 \mathsf C_{c=1}((XHX)_t).
\end{equation}
After all selection rotations, process the expansion stages
in increasing order $r=0,1,\ldots,h-2$. At stage $r$,
for each $t=e_r,\ldots,e_{r+1}-1$, apply
$\C{q_{e_{r+1}}=1}{(XHX)_{q_t}}$.
The control selects precisely the reflected blocks $D_s$
with $s\geq r+1$. On these blocks, the target qubits are
in $\ket1$, so each controlled operation replaces the
target state by $\ket+$. The already selected blocks
$D_s$ with $s\leq r$ are unaffected. Thus, the reflected
expansion is the counterpart of the controlled-Hadamard
expansion in Lemma~\ref{lem:common-partition}, with every
operation remaining a singly controlled one-qubit gate.

To identify the resulting state, consider a branch of length
$m$. At stage $r$, the active component has squared norm
$V_r(m)/m$, while the component selected into $D_r$ has
squared norm $P_r(m)/m$. For a selected block $D_r$, the
expansion gates act on the ranges $[0,v)$ and
$[e_s,e_{s+1})$, $0\leq s<r$. These ranges partition
$[0,e_r)$, so exactly $e_r$ qubits are put into the state
$\ket+$. Each basis state in the selected block therefore
has amplitude
$\sqrt{2^{e_r}/m}\,2^{-e_r/2}=1/\sqrt m$.
Since the selected reflected blocks are disjoint and
$b_{h-1-r}(m)$ records whether $D_r$ is selected, the
lower-register output before the final branch masks is
\begin{equation}\label{eq:reflected-state}
 \ket{\widetilde\Phi_m^S}
 =
 \frac1{\sqrt m}
 \sum_{\substack{0\leq r<h\\b_{h-1-r}(m)=1}}
 \sum_{x\in D_r}\ket{x}
 =
 X^{F_{\mathrm{act}}}\ket{\Phi_m^S}.
\end{equation}
For $h=1$, the same identity follows directly from the
initial bit flips and Hadamard gates.

\paragraph{Final coordinate corrections.}
The active reflection need not complement the entire
$k$-bit register. To account for the unchanged high padding
qubits, define
$P_{\mathrm{pad}}=\sum_{t=\lambda_0+1}^{k-1}2^t$,
where the sum is zero when $\lambda_0=k-1$. With
$F=2^k-1$, we have
$F_{\mathrm{act}}=F\oplus P_{\mathrm{pad}}$.
The final branch masks are
\begin{equation}\label{eq:reflected-final-masks}
 \tau_L^{(0)}=\sigma_L\oplus P_{\mathrm{pad}},
 \qquad
 \tau_R^{(0)}=F\oplus\sigma_R\oplus P_{\mathrm{pad}}.
\end{equation}
By Lemma~\ref{lem:xor-conversion}, these masks give
\begin{align}
 X^{\tau_L^{(0)}}\ket{\widetilde\Phi_L^S}
 &=X^F X^{\sigma_L}\ket{\Phi_L^S}
 =X^F\ket{\psi_L^{(k)}},\\
 X^{\tau_R^{(0)}}\ket{\widetilde\Phi_R^S}
 &=X^{\sigma_R}\ket{\Phi_R^S}
 =\ket{\psi_R^{(k)}}.
\end{align}

As in the direct construction, the lowest $\nu_2(m)$
mask bits may be omitted because the corresponding qubits
are already in $\ket+$. The reduced masks are
$\widehat{\tau}_L^{(0)}
=\tau_L^{(0)}\mathbin{\land}(2^k-2^{\nu_2(L)})$
and
$\widehat{\tau}_R^{(0)}
=\tau_R^{(0)}\mathbin{\land}(2^k-2^{\nu_2(R)})$.
Implement them as in \Cref{alg:interval-main}: a bit present in both
masks requires an unconditional $X$; a bit present only
in the lower mask requires an $X$ controlled on $q_k=0$;
and a bit present only in the upper mask requires an $X$
controlled on $q_k=1$. Finally, set the common high-order
prefix $Q$ using Pauli-$X$ gates on its set-bit positions.
The resulting state is exactly the desired uniform
superposition over $[a,a+M)$.

The reflected construction therefore uses only the $n$
data qubits and one-qubit or singly controlled one-qubit
gates, without ancillary qubits.

\subsection{Gate-count comparison for aligned suffixes}

We record the comparison with the direct construction
for an aligned suffix. Suppose $2^k<M<2^{k+1}$ and write
$M=2^{v_M}N_M$, where
$s_M=\operatorname{wt}(N_M)$ and $h_M=k-v_M+1$.
The direct complement-adapted realization of
Lemma~\ref{lem:aligned-suffix} uses
$C_1=K+h_M+s_M-3$ logical gates, as established in
\eqref{eq:direct-suffix-C1}, with $K=k+1$.

In the reflected construction, all blocks are selected.
The initial bit flips contribute $h_M-s_M+1$ gates,
the Hadamards contribute $v_M$, the block-selection
rotations contribute $s_M-1$, and the controlled $XHX$
expansions contribute $h_M-1$. Therefore,
\begin{equation}\label{eq:reflected-C0}
\begin{aligned}
 C_0
 =(h_M-s_M+1)+v_M+(s_M-1)+(h_M-1)
 =2h_M+v_M-1
 =K+h_M-1.
\end{aligned}
\end{equation}
It follows that
\begin{equation}\label{eq:reflected-count-difference}
 C_1-C_0=s_M-2\geq0.
\end{equation}
Thus, the reflected suffix circuit is never more expensive:
the two constructions have equal counts when $s_M=2$,
and the reflected construction is strictly shorter when
$s_M\geq3$. The common high-order prefix contributes the
same additional $\operatorname{wt}(Q)$ gates to both.
The full dyadic block $M=2^{k+1}$ is prepared directly
using $K$ Hadamard gates.

For general intervals, the two constructions can also
have substantially different gate counts. For example,
on $[1,65)$ in a seven-qubit register, the direct
construction uses 26 logical gates, whereas the reflected
construction uses 13. This illustrates the instance-dependent
advantage of reversing the block processing order. The
classical selection of the lower-count complete circuit is
described in \Cref{subsec:gate-count-adaptive}.

\section{XOR images and arithmetic progressions}\label{sec:xor-progressions}

The interval circuit has two extensions to supports that need not be contiguous. The first is closure under bitwise XOR. The second embeds an interval in a fixed residue class modulo a power of two by placing the interval register in the higher-order qubits.

\subsection{XOR images of intervals}

For an $n$-bit mask $s$, the operator defined in
\Cref{sec:setting} satisfies $X^s\ket{x}=\ket{x\oplus s}$.
We obtain the following immediate consequence.

\begin{proposition}[Uniform superpositions on XOR images]\label{prop:xor-image}
For every valid interval $[a,a+M)$ and every $n$-bit mask $s$,
\[
 X^s\ket{\psi_{a,M}^{(n)}}
 =\frac1{\sqrt M}\sum_{x=a}^{a+M-1}\ket{x\oplus s}.
\]
The state is a uniform superposition over the set
$ [a,a+M)\oplus s
 :=\{x\oplus s:a\leq x<a+M\}$.
It is prepared exactly by appending $\wt(s)$ one-qubit Pauli-$X$ gates to the interval-preparation circuit. No ancillary qubit or controlled gate is required for this extension.
\end{proposition}

\begin{proof}
Bitwise XOR by a fixed mask is a bijection on $n$-bit strings. Thus distinct values of $x$ in the interval have distinct images, and applying $X^s$ maps every basis vector in the interval state to its corresponding XOR image without changing its amplitude. The mask contains $\wt(s)$ set bits, so exactly that many Pauli-$X$ gates suffice.
\end{proof}

The XOR image need not be a contiguous interval or an arithmetic progression. It nevertheless has a dyadic description. A dyadic block of size $2^r$ has the form
 $B(u,r)=[u2^r,(u+1)2^r)$.
Writing $s=s_{\mathrm{hi}}2^r+s_{\mathrm{lo}}$ with $0\leq s_{\mathrm{lo}}<2^r$, one has $ B(u,r)\oplus s=B\bigl(u\oplus s_{\mathrm{hi}},r\bigr)$,
because XOR by $s_{\mathrm{lo}}$ permutes all $r$-bit strings within the block. Since every integer interval can be partitioned into disjoint aligned dyadic blocks, its XOR image is a disjoint union of dyadic blocks of the same respective sizes. The circuit preparation itself does not need to construct this decomposition: Proposition~\ref{prop:xor-image} applies the mask directly.

\begin{remark}[XOR versus integer translation]
The operation $X^s$ implements XOR, not addition. It gives an ordinary integer translation by $s$ on an interval $I=[a,a+M)$ precisely when
\[
 s\land\left(\bigvee_{x=a}^{a+M-1}x\right)=0,
 \qquad a+M-1+s<2^n,
\]
where $\bigvee$ denotes bitwise OR over the binary representations of the integers in the interval. The first condition says that no integer in the support has a set bit at a position where $s$ also has a set bit, so the addition produces no carries. Under these conditions $x+s=x\oplus s$ for every $x\in I$, and $X^s$ prepares the uniform state on $[a+s,a+M+s)$. For a shift by $2^t$, it is sufficient and necessary that bit $t$ be zero on every occupied basis state, together with the no-overflow condition.
\end{remark}

\subsection{Arithmetic progressions with power-of-two spacing}

Let $0\leq r<n$, let $0\leq c<2^r$, and suppose $0\leq a<a+M\leq2^{n-r}$. Define
\begin{equation}\label{eq:fixed-residue-progression-state}
 \ket{\psi_{a,M}^{(n;r,c)}}
 =\frac1{\sqrt M}\sum_{j=a}^{a+M-1}\ket{2^rj+c}.
\end{equation}
This is a uniform superposition over an arithmetic progression with common difference $2^r$ and residue $c$ modulo $2^r$.

\begin{theorem}[Ancilla-free preparation of fixed-residue progressions]
\label{thm:fixed-residue-progressions}
Let $0\leq r<n$, let $c$ be an integer satisfying
$0\leq c<2^r$, and let $a,M$ be integers satisfying
    $0\leq a<a+M\leq 2^{n-r}$.
The uniform state
\begin{equation*}
    \ket{\psi_{a,M}^{(n;r,c)}}
    =
    \frac{1}{\sqrt M}
    \sum_{j=a}^{a+M-1}\ket{2^rj+c}
\end{equation*}
can be prepared by applying the common dyadic-block interval
circuit for $[a,a+M)$ to the upper $n-r$ qubits
$q_{n-1},\ldots,q_r$, leaving the lower $r$ qubits in
$\ket{0^r}$, and then applying $X$ to the lower qubits
corresponding to the set bits of $c$. The construction is
exact and ancilla-free. Its logical gate count is at most
\[
    G_{\mathrm{dir}}^{(n-r)}(a,M)+\operatorname{wt}(c),
\]
where $G_{\mathrm{dir}}^{(n-r)}(a,M)$ denotes the logical
gate count of the direct interval-preparation circuit on
the upper register of $n-r$ qubits.
\end{theorem}

\begin{proof}
We note that each computational-basis state
factors as
 $\ket{2^rj+c}
 =
 \ket{j}_{q_{n-1}\cdots q_r}
 \otimes
 \ket{c}_{q_{r-1}\cdots q_0}$.
Therefore, the desired state has the tensor-product
decomposition
\[
\begin{aligned}
 \ket{\psi_{a,M}^{(n;r,c)}}
 =
 \left(
 \frac{1}{\sqrt M}
 \sum_{j=a}^{a+M-1}
 \ket{j}_{q_{n-1}\cdots q_r}
 \right)
 \otimes
 \ket{c}_{q_{r-1}\cdots q_0}
 =
 \ket{\psi_{a,M}^{(n-r)}}\otimes\ket{c}.
\end{aligned}
\]
The interval-preparation circuit prepares
$\ket{\psi_{a,M}^{(n-r)}}$ on the upper register, while
the lower register remains in $\ket{0^r}$. Applying $X$
to each lower qubit corresponding to a set bit of $c$
then prepares $\ket{c}$. The resulting product state is
exactly the state in \eqref{eq:fixed-residue-progression-state}.

The upper-register preparation requires
$G_{\mathrm{dir}}^{(n-r)}(a,M)$ logical gates, and the
lower-register preparation requires $\operatorname{wt}(c)$
Pauli-$X$ gates. Since these operations act on disjoint
registers, their gate counts add. The construction uses
only the $n$ data qubits and is therefore ancilla-free.
\end{proof}

For $r=1$, $c=0$ and $c=1$ give the even- and odd-indexed states, respectively:
\[
 \ket{\psi_{a,M}^{(n;1,0)}}=\frac1{\sqrt M}\sum_{j=a}^{a+M-1}\ket{2j},
 \qquad
 \ket{\psi_{a,M}^{(n;1,1)}}=\frac1{\sqrt M}\sum_{j=a}^{a+M-1}\ket{2j+1}.
\]
For example, with $a=1$ and $M=4$, these are
$\frac12(\ket2+\ket4+\ket6+\ket8)$ and
$\frac12(\ket3+\ket5+\ket7+\ket9)$.
The same construction gives every residue class modulo $2^r$, not only the even and odd cases.

More generally, let $C\subseteq\{0,\ldots,2^r-1\}$ be
a nonempty set of cardinality $K$, and suppose
\[
    \ket{\chi_C}
    =\frac1{\sqrt K}\sum_{c\in C}\ket c
\]
can be prepared from $\ket{0^r}$ by a circuit acting
only on the lower $r$ qubits, using
$G_{\mathrm{low}}(C)$ logical gates. Then
\[
 \ket{\psi_{a,M}^{(n-r)}}_{\mathrm{high}}
 \otimes\ket{\chi_C}_{\mathrm{low}}
 =
 \frac1{\sqrt{KM}}
 \sum_{c\in C}\sum_{j=a}^{a+M-1}
 \ket{2^rj+c}
\]
is a uniform superposition over the union of $K$
arithmetic progressions with common difference $2^r$.
It uses at most
\[
 G_{\mathrm{dir}}^{(n-r)}(a,M)
 +G_{\mathrm{low}}(C)
\]
logical gates. The construction is ancilla-free whenever
the chosen preparation circuit for $\ket{\chi_C}$
is itself ancilla-free.

\section{Gate complexity and depth}\label{sec:complexity}

We count each one-qubit gate and each singly controlled one-qubit gate as one logical gate. Positive and negative controls are both allowed, and a controlled-$Z$ operation is included among the singly controlled one-qubit gates. At a noninitial block-selection stage, the three-gate paired-angle identity of Lemma~\ref{lem:paired-angle} implements both branch rotations on a target known to be in $\ket0$. It is a restricted state-preparation identity, not a general two-control unitary decomposition. Arbitrary real rotation angles are allowed. This is a logical gate count, not a CNOT count after decomposition.

\begin{proposition}[Uncorrected gate count of the common dyadic-block partition]\label{prop:partition-count}
For $S=L\mathbin{\lor}R$ with binary positions $\lambda_0>\cdots>\lambda_{h-1}=v$, the common gate pattern before branch-dependent angle corrections has baseline count
\begin{equation}\label{eq:partition-count}
  v+(h-1)+(\lambda_0-v)=\lambda_0+h-1
\end{equation}
logical gates: $v$ initial Hadamards, $h-1$ baseline rotations, and $\lambda_0-v$ controlled Hadamards. The actual circuit count also includes any branch-dependent rotation corrections, which are accounted for in Theorem~\ref{thm:interval-upper}.
\end{proposition}

\begin{proof}
The initial low-order Hadamards contribute $v$. The rotation sequence has one rotation at stage $j=0$ and one controlled rotation at each subsequent stage $1,\ldots,h-2$, for $h-1$ baseline rotations. The controlled-Hadamard targets are the disjoint ranges $[\lambda_{j+1},\lambda_j)$, so their number is
\[
 \sum_{j=0}^{h-2}(\lambda_j-\lambda_{j+1})=\lambda_0-v.
\]
Adding these contributions gives \eqref{eq:partition-count}. If a baseline angle is zero, its identity rotation may be omitted; hence \eqref{eq:partition-count} is an upper bound for the circuit implementing the common dyadic-block partition and is attained whenever all baseline rotations are nonzero.
\end{proof}

\begin{theorem}[Exact worst-case logical gate count of the direct construction]
\label{thm:interval-upper}

Let $G_n$ be the maximum logical gate count of
Algorithm~\ref{alg:interval-main}, which is the direct
construction, over all nonempty intervals in an $n$-qubit
register. Then
\[
  G_1=1,\qquad G_2=3,\qquad G_n=5n-9\quad(n\geq3).
\]
Thus, $5n-9$ is the exact worst-case gate count of the
specified direct construction. In particular, the worst-case
gate complexity is $\Theta(n)$.
\end{theorem}

\begin{proof}
Singleton intervals require at most $n$ Pauli-$X$ gates. Consider the non-boundary case $1\leq L,R<2^k$ with $h\geq2$. Let $\delta_0$ equal one if both stage-zero angles are nonzero and unequal, and zero otherwise. Let $r$ be the number of stages $1\leq j\leq h-2$ at which both branches remain active and their angles differ. The common dyadic-block partition contributes at most $\lambda_0+h-1$ gates. The selector rotation contributes one gate. At stage zero, at most $\delta_0$ additional gate is required beyond the single baseline operation. At every later corrected stage, the three-gate paired-angle circuit replaces one baseline controlled rotation, adding two gates. The final reduced masks require $\operatorname{wt}(\widehat\tau_L\mathbin{\lor}\widehat\tau_R)$ gates, and the common high-order prefix requires $\operatorname{wt}(Q)$ gates. Therefore
\begin{equation}\label{eq:partition-count-upper}
 G\leq \lambda_0+h+\delta_0+2r
       +\operatorname{wt}(\widehat\tau_L\mathbin{\lor}\widehat\tau_R)
       +\operatorname{wt}(Q).
\end{equation}
Here $\lambda_0\leq k-1$, $h\leq k$, $\delta_0\leq1$, $r\leq h-2\leq k-2$, $\operatorname{wt}(\widehat\tau_L\mathbin{\lor}\widehat\tau_R)\leq k$, and $\operatorname{wt}(Q)\leq n-k-1$. Therefore, 
 $G\leq(k-1)+k+1+2(k-2)+k+(n-k-1) =4k+n-5 
  \leq5n-9$,
where $k\leq n-1$.

If $h=1$, then $L=R=S=2^v$. The initial Hadamards and selector
rotation contribute $v+1$ gates, while the final branch mask
contributes $k-v$ controlled bit flips. Including the
$\operatorname{wt}(Q)$ gates for the common high-order prefix,
the total is
\[
v+1+(k-v)+\operatorname{wt}(Q)
=k+1+\operatorname{wt}(Q)\leq n.
\]
For a non-full aligned prefix, $L=2^k$ and $0<R<2^k$, so $M=2^k+R<2^{k+1}$. Write $M=2^{v_M}N_M$ with $N_M$ odd, and let $s_M=\wt(N_M)$. Algorithm~\ref{alg:prefix-state} uses $s_M-1$ initial bit flips, $v_M$ Hadamards, $s_M-1$ rotations, and $k-v_M$ controlled Hadamards. Its count on the lower $k+1$ qubits is $k+2s_M-2$. Since $s_M\leq k-v_M+1$, adding at most $n-k-1$ gates for the common high-order prefix gives a total no greater than $n+2k-2v_M-1\leq3n-3$. For a non-full aligned suffix, $R=2^k$ and $0<L<2^k$. The complement-reduced direct suffix circuit uses at most $(2k-v_M)+(k-v_M)$ gates on the lower register; including the at most $n-k-1$ high-prefix flips gives at most $n+2k-2v_M-1\leq3n-3$. The full parent block $L=R=2^k$ requires $k+1$ Hadamards on the lower register and at most $n-k-1$ high-prefix flips, for a total of at most $n$. All these boundary counts are at most $5n-9$ for $n\geq3$. Direct inspection gives $G_1=1$ and $G_2=3$.

We finally show that the upper bound is attained for every
$n\geq3$. Put $k=n-1$ and consider the interval $ I_n=[1,2^k+1)$.
Its midpoint-split parameters are
    $L=2^k-1$, $R=1$, and $Q=0$,
and hence $ S=L\mathbin{\lor}R=2^k-1$.
Therefore,
$ h=k$, $\lambda_j=k-1-j$, $(0\leq j\leq k-1)$ and $v=0$.
At every selection stage $0\leq j\leq k-2$, both branches
remain active and their angles differ. More precisely,
\[
\theta_j(L)
=
2\arccos\sqrt{
\frac{2^{k-j-1}}{2^{k-j}-1}},
\qquad
\theta_j(R)=\pi.
\]
Thus the first stage requires two rotations, while each
of the subsequent $k-2$ stages requires the three-gate
paired-angle construction.
The final reduced masks are
    $\widehat{\tau}_L=2^k-1$ and
    $\widehat{\tau}_R=2^k-2$.
Their bitwise OR has Hamming weight $k$. The controlled-
Hadamard expansion uses $k-1$ gates. Including the selector
rotation, the complete direct gate count is therefore
\[
\begin{aligned}
G_{\mathrm{dir}}(1,2^k)
 =1+2+3(k-2)+(k-1)+k
 =5k-4
 =5n-9.
\end{aligned}
\]
This proves that the upper bound is attained for every
$n\geq3$, completing the proof.
\end{proof}

\begin{proposition}[Instance-dependent logical gate bound]
\label{prop:instance-dependent-bound}
Define the offset-dependent parameter $\epsilon(a,M)$ as follows.
For a singleton interval, put
$\epsilon(a,1)=\operatorname{wt}(a)$.
For $M>1$, use the midpoint-split parameters
$k,L,R,Q$ from \eqref{eq:midpoint-common}, and set
\[
\epsilon(a,M)=
\begin{cases}
\operatorname{wt}(Q)+
\operatorname{wt}(\widehat{\tau}_L\mathbin{\lor}
                   \widehat{\tau}_R),
 & 0<L,R<2^k,\\[1mm]
\operatorname{wt}(Q),
 & L=2^k\text{ or }R=2^k.
\end{cases}
\]
Then the logical gate count of the direct construction satisfies
\[
G_{\mathrm{dir}}(a,M)
\leq
4\lfloor\log_2 M\rfloor+\epsilon(a,M),
\qquad M>1.
\]
For a singleton interval, the exact count is
$G_{\mathrm{dir}}(a,1)=\epsilon(a,1)=\operatorname{wt}(a)$.
Therefore, for all nonempty intervals,
\[
G_{\mathrm{dir}}(a,M)
=O\bigl(\log_2 M+\epsilon(a,M)\bigr).
\]
The adaptive compiler satisfies the same instance-dependent
asymptotic bound.
\end{proposition}

The proof of this proposition is presented in Appendix~\ref{app:additional-complexity-proofs}.

\begin{theorem}[Worst-case linear lower bound]\label{thm:interval-lower}
In a circuit model in which every gate acts on at most two qubits, the worst-case gate complexity of preparing $\ket{\psi_{a,M}^{(n)}}$ over all allowed intervals is $\Omega(n)$. Therefore, Algorithm~\ref{alg:interval-main} is asymptotically optimal in register width.
\end{theorem}

\begin{proof}
Consider the full interval $[0,2^n)$, whose target is $\ket{+}^{\otimes n}$. Every output qubit must be acted upon by at least one gate, since an untouched qubit remains in $\ket0$ and cannot have reduced state $\ket+\bra+$. A gate acts on at most two qubits, so at least $\lceil n/2\rceil$ gates are required. This is an $\Omega(n)$ lower bound. Together with Theorem~\ref{thm:interval-upper}, it establishes $\Theta(n)$ worst-case complexity.
\end{proof}

\subsection{Logical circuit depth}
\label{subsec:logical-depth}

We use the standard logical-depth convention in which a layer contains gates acting on pairwise disjoint sets of qubits. A singly controlled one-qubit gate occupies both its control and target in that layer. The depth model uses the same logical gate alphabet as the gate-count analysis: arbitrary one-qubit gates and singly controlled one-qubit gates, with either control polarity. We assume unrestricted connectivity and do not include the additional depth required to decompose controlled gates into a hardware-specific gate set or to route them on a restricted architecture. Let $D(a,M)$ denote the minimum number of such layers required to schedule the logical circuit specified by Algorithm~\ref{alg:interval-main} for the interval $[a,a+M)$.

The same stage structure that gives the linear gate bound also gives a linear depth upper bound. The initial selector rotation, the low-order Hadamards, and the flips that set the common high-order prefix act on disjoint qubits and can be placed in one layer. At stage zero, at most two rotations act on the same target, so at most two further layers are needed. Each later paired-angle stage uses at most three singly controlled gates. The controlled-Hadamard expansion contains $\lambda_0-v$ gates in total; a conservative schedule places them in separate layers. Finally, the branch-dependent mask acts on the $k$ lower qubits. Its controlled flips share the selector and are therefore scheduled one at a time, while the unconditional flips can be run in parallel with this sequence. Thus the final branch mask requires at most $k$ additional layers. The high-prefix flips are already included in the initialization layer, since they act on wires disjoint from all lower-register operations.

\begin{proposition}[Logical-depth upper bound]\label{prop:depth-upper}
For a non-boundary interval with $1\leq L,R<2^k$ and $h=\operatorname{wt}(L\mathbin{\lor}R)\geq2$, the direct construction admits a schedule of depth
\begin{equation}\label{eq:generic-depth}
 D(a,M)\leq 1+2+3(h-2)+(\lambda_0-v)+k
 =\lambda_0-v+3h+k-3.
\end{equation}
Therefore, for every nonempty interval in an $n$-qubit register,
\[
 D_1\leq1,\qquad D_2\leq3,\qquad D_n\leq5n-9\quad(n\geq3),
\]
where $D_n$ denotes the maximum depth of Algorithm~\ref{alg:interval-main} over all such intervals.
\end{proposition}

The proof of this proposition is given in Appendix~\ref{app:additional-complexity-proofs}.

The depth estimate is an upper bound for the stated logical model, not a hardware-depth prediction. In particular, controlled-Hadamard gates within an expansion stage share a control qubit, and the branch-dependent controlled flips share the selector; these dependencies limit parallelization in the present ancilla-free construction. We do not claim a matching $\Omega(n)$ depth lower bound: the gate-count lower bound in Theorem~\ref{thm:interval-lower} does not imply a depth lower bound because gates on disjoint qubits may act in parallel.

\begin{remark}
For a particular interval, the count can be substantially smaller than the worst-case bound. In particular, when $L=R$, all branch-dependent rotation corrections vanish. The circuit then reduces to the ordinary dyadic-block prefix preparation for $L$, followed by a Hadamard on the selector and at most $k-v_L$ selector-controlled bit flips, where $v_L=\nu_2(L)$; the lowest $v_L$ flips are redundant because those qubits are already in $\ket+$. No ancillary qubit or branch-dependent rotation is required.
\end{remark}

\subsection{Classical selection of the lower-count circuit}
\label{subsec:gate-count-adaptive}

The direct and reflected constructions prepare the same target
state, but their complete logical gate counts can differ
substantially from one interval to another. We therefore use
a classical compiler to select the less expensive exact
realization. The compiler constructs the applicable circuit
descriptions, counts their logical gates, and returns one
circuit for execution; no coherent selection register is
required.

The candidate set depends on the position of the interval
relative to the midpoint split. For a generic interval,
$1\leq L,R<2^k$, let $\mathcal C_1(a,M)$ and
$\mathcal C_0(a,M)$ denote the complete direct and reflected
circuits, respectively. For an aligned suffix,
$0<L<2^k$ and $R=2^k$, the compiler uses the reflected
suffix circuit $\mathcal C_0^{\mathrm{suffix}}(a,M)$
described in \Cref{subsec:reflected-suffix}.

This choice is justified by
\eqref{eq:reflected-count-difference}: the reflected suffix
circuit is never more expensive than the direct
complement-adapted suffix circuit. The common high-order
prefix contributes the same additional
$\operatorname{wt}(Q)$ gates to both. Therefore, no
comparison of two suffix candidates is necessary.

For a generic interval, write $G_1(a,M)$ and $G_0(a,M)$
for the complete logical gate counts of the direct and
reflected circuits. For an aligned suffix, let
$G_0^{\mathrm{suffix}}(a,M)$ denote the complete count
of the reflected suffix circuit, including the common
high-order prefix. The adaptive gate count is
\begin{equation}\label{eq:adaptive-gate-count}
G_{\mathrm{ad}}(a,M)=
\begin{cases}
\min\{G_1(a,M),G_0(a,M)\},
  & 1\leq L,R<2^k,\\
G_0^{\mathrm{suffix}}(a,M),
  & 0<L<2^k,\ R=2^k,\\
G_{\mathrm{sp}}(a,M),
  & M=1\text{ or }L=2^k.
\end{cases}
\end{equation}
Here $G_{\mathrm{sp}}$ is the count of the appropriate
specialized routine: the singleton circuit when $M=1$,
and the aligned-prefix circuit when $L=2^k$, including
the full parent block. These cases are disjoint and
cover every nonempty interval. The alignment conditions
refer to the smallest dyadic block that can contain an
interval of length $M$, as specified in
\Cref{subsec:special-cases}.

\begin{algorithm}[H]
\caption{Classical selection of the lower-count circuit
for interval preparation}
\label{alg:adaptive}
\begin{algorithmic}[1]
\Require Register width $n$ and a valid interval
         $[a,a+M)$
\If{$M=1$}
    \State Construct and \Return the singleton circuit
\Else
    \State Compute the midpoint split $(k,L,R,Q)$
    \If{$L=2^k$}
        \State Construct and \Return the aligned-prefix circuit
    \ElsIf{$R=2^k$}
        \State Construct the reflected suffix circuit
               $\mathcal C_0^{\mathrm{suffix}}(a,M)$
        \State \Return $\mathcal C_0^{\mathrm{suffix}}(a,M)$
    \Else
        \State Construct the complete direct circuit
               $\mathcal C_1(a,M)$
        \State Construct the complete reflected circuit
               $\mathcal C_0(a,M)$
        \State Compute their complete logical gate counts
        \State \Return the candidate with smaller count
    \EndIf
\EndIf
\end{algorithmic}
\end{algorithm}

\Needspace{10\baselineskip}
\begin{theorem}[Correctness and worst-case complexity
of the adaptive compiler]
\label{thm:adaptive-compiler}

For every nonempty interval $[a,a+M)$ in an $n$-qubit
register, Algorithm~\ref{alg:adaptive} returns an exact,
ancilla-free circuit using only one-qubit and singly
controlled one-qubit gates. Let $G_{\mathrm{ad},n}$ be
the maximum logical gate count of the returned circuit
over all nonempty intervals in an $n$-qubit register.
Then $G_{\mathrm{ad},1}=1$, $G_{\mathrm{ad},2}=3$, and
$G_{\mathrm{ad},n}\leq5n-9$ for $n\geq3$.
Moreover, the worst-case gate complexity of the adaptive
implementation is $\Theta(n)$.
\end{theorem}

\begin{proof}
Every candidate circuit is exact and ancilla-free by the
direct construction, the reflected construction in
\Cref{sec:reflected-construction}, or the aligned-suffix
specialization in \Cref{subsec:reflected-suffix}.
The compiler returns one complete candidate, so its
output has the required properties.

For a generic interval, the compiler takes the minimum
of the complete direct and reflected counts and hence
cannot use more gates than the direct construction.
For an aligned suffix, the same conclusion follows from
\eqref{eq:reflected-count-difference}, since the reflected
suffix circuit is no more expensive than the direct
complement-adapted suffix circuit. For a singleton or
aligned-prefix interval, the compiler uses the
corresponding direct specialized circuit. Thus, for
every interval,
\[
G_{\mathrm{ad}}(a,M)\leq G_1(a,M),
\]
where $G_1(a,M)$ denotes the complete count of the main
direct algorithm, including its specialized cases.
Theorem~\ref{thm:interval-upper} therefore gives the
stated upper bounds.

For $n=1$, the interval $[1,2)$ requires one Pauli-$X$
gate, so the maximum is exactly one. For $n=2$, the
interval $[1,4)$ is an aligned suffix. Its reflected
suffix circuit uses three logical gates. Since the
adaptive count is bounded above by the direct
construction's exact worst-case count of three, the
maximum is exactly three.

For the lower bound, consider the full interval
$[0,2^n)$, whose target is $\ket{+}^{\otimes n}$.
As in Theorem~\ref{thm:interval-lower}, every output
qubit must be acted upon by a gate, and a gate acts on
at most two qubits. Thus any implementation requires
at least $\lceil n/2\rceil$ gates. This gives an
$\Omega(n)$ worst-case lower bound, which together
with the upper bound proves $\Theta(n)$.
\end{proof}

The classical selection overhead is linear in the register
width when bit operations on the binary interval data are
counted individually. For a generic interval, both
candidate circuits have $O(n)$ stages and logical
operations; their masks and stage parameters are
determined by scans of $O(n)$-bit binary representations.
For aligned suffixes, only the reflected candidate is
constructed. The specialized prefix and singleton cases
also require only $O(n)$ classical work. Thus, the
complete compiler requires $O(n)$ classical bit-level
work, in addition to evaluating the rotation angles.
No quantum operation is added for the comparison itself.

Let $D_{\mathrm{ad}}(a,M)$ denote the logical depth of the
circuit returned by Algorithm~\ref{alg:adaptive}, under the
layer convention of \Cref{subsec:logical-depth}. Let
$D_{\mathrm{ad},n}$ be the maximum of this depth over all
nonempty intervals in an $n$-qubit register.

The adaptive compiler minimizes logical gate count and need
not select the circuit with smaller logical depth for a
particular interval. Nevertheless, its selected circuit
satisfies
\[
  D_{\mathrm{ad}}(a,M)
  \leq G_{\mathrm{ad}}(a,M)
  \leq G_{\mathrm{dir}}(a,M).
\]
The first inequality follows because any circuit can be
scheduled serially, with one gate per layer. The second
inequality follows from
Proposition~\ref{prop:instance-dependent-bound}.
Consequently, the adaptive construction inherits the global
depth bounds
\[
  D_{\mathrm{ad},1}\leq1,\qquad
  D_{\mathrm{ad},2}\leq3,\qquad
  D_{\mathrm{ad},n}\leq5n-9
  \quad(n\geq3).
\]

The more detailed parameter-dependent schedule bound in
Proposition~\ref{prop:depth-upper} is established for the
direct construction. We do not claim that the adaptive
compiler minimizes logical depth for individual intervals.
A  compiler could be designed to additionally compare the
depths of the candidate circuits and optimize the circuit based on the depth count.

\section{Examples}
\label{sec:examples}

\subsection{Equal branch lengths: the interval
\texorpdfstring{$[3,13)$}{[3,13)}}

Take $n=4$, $a=3$, and $M=10$. The endpoints are $3$ and
$12$, so $k=3$, $D=8$, and $L=R=5$. The common high-order
bit string is zero. Since the branch lengths agree, all
branch-dependent rotation corrections disappear. The circuit is:
\begin{enumerate}[leftmargin=2em]
  \item Prepare $\ket{\psi_5^{(3)}}$ on $q_2q_1q_0$ using
  the common-partition block routine for $5=4+1$.
  \item Apply $H$ to the selector $q_3$.
  \item Apply $X$ to each lower qubit controlled on $q_3=0$.
\end{enumerate}
The output is
$\frac1{\sqrt{10}}\left(\ket3+\ket4+\cdots+\ket{12}\right)$.
This block routine uses three logical gates (one rotation
and two controlled Hadamards). The selector rotation and
the three selector-controlled bit flips require four more
gates, for a total of seven logical gates.

\subsection{A complete circuit derivation: the interval
\texorpdfstring{$[3,11)$}{[3,11)}}
\label{subsec:worked-example}

We illustrate the construction in detail for the four-qubit
interval $I=[3,11)=\{3,4,5,6,7,8,9,10\}$. This example is
useful because neither branch is aligned with a dyadic
boundary, and both branches require a nontrivial
block-selection rotation correction. It therefore illustrates
all the main ingredients of the construction: splitting the
interval at a dyadic midpoint, selecting dyadic blocks,
expanding the selected blocks, and finally remapping them
by branch-dependent XOR masks. We use little-endian indexing,
writing a computational-basis state as $\ket{q_3q_2q_1q_0}$.

\paragraph{Step 1: Split the interval at its dyadic midpoint.}

Here $n=4$, $a=3$, $M=8$, and $b=a+M-1=10$. In binary,
$a=0011_2$, $b=1010_2$, and
$a\mathbin{\oplus}b=1001_2$. The most significant differing
bit is $k=3$. Thus the dyadic parent block has midpoint
$D=8$, and the interval splits into
$[3,11)=[3,8)\cup[8,11)$. The two branch lengths are
$L=D-a=5$ and $R=a+M-D=3$.

The most significant qubit $q_3$ serves as the branch selector:
$q_3=0$ identifies the lower branch $[3,8)$, while
$q_3=1$ identifies the upper branch $[8,11)$. There is no
common high-order prefix to prepare in this example.

\paragraph{Step 2: Construct a common dyadic-block partition.}

The two branch lengths have different binary decompositions:
$L=5=4+1$ and $R=3=2+1$. Two branches are not constructed independently. Instead, we use their bitwise OR,
$S=L\mathbin{\lor}R=5\mathbin{\lor}3=7=2^2+2^1+2^0$.
This gives a common sequence of dyadic block sizes,
$(\lambda_0,\lambda_1,\lambda_2)=(2,1,0)$, with
$h=3$ and $v=0$. The corresponding blocks are
$B_0=[0,4)$, $B_1=[4,6)$, and $B_2=[6,7)$.

The key idea is that both branches use this same partition.
The lower branch selects $B_0$ and $B_2$, since $5=4+1$,
whereas the upper branch selects $B_1$ and $B_2$, since
$3=2+1$. The block-selection qubits are therefore the same
for both branches: $q_2$ selects the block of size four,
and $q_1$ selects the block of size two. The rotations
determine which blocks are populated on each branch.

The selection process can be understood by keeping track
of the remaining length at each stage. When a block of size
$2^{\lambda_j}$ is present, we select it and subtract its
size from the residual length. Otherwise, we skip that
block and retain the entire residual for the next stage.
For this example, the residual lengths and block-selection
indicators are

\[
\begin{array}{c|c|cc|cc}
\text{stage }j & 2^{\lambda_j}
 & r_j(L) & b_j(L) & r_j(R) & b_j(R)\\\hline
0&4&5&1&3&0\\
1&2&1&0&3&1\\
2&1&1&1&1&1
\end{array}
\]

Here $b_j(m)=1$ means that the block of size $2^{\lambda_j}$
is selected for branch $m$, whereas $b_j(m)=0$ means that
it is skipped. The residual length is reduced by the block
size precisely when the corresponding indicator is one.

Thus, on the lower branch, the size-four block is selected
first, leaving a residual length of one. The size-two block
is then skipped, and the final singleton block is selected.
On the upper branch, the size-four block is skipped, the
size-two block is selected, and the singleton block is
selected as well.

The rotation angles implement these choices while assigning
probability in proportion to block size. If the current
residual length is $r$ and a block of size $d$ is selected,
the rotation is chosen so that the probability of selecting
that block is $d/r$. If the block is absent, the angle is
$\pi$, which transfers the entire amplitude to the
continuation. If the residual equals the current block size,
no rotation is needed: the block is selected with probability
one.

For the present example, the required angles are

\[
\begin{array}{c|cc}
 &\text{lower branch}&\text{upper branch}\\\hline
j=0&2\arccos\sqrt{\frac45}&\pi\\[1mm]
j=1&\pi&2\arccos\sqrt{\frac23}\\[1mm]
j=2&0&0
\end{array}
\]

Equivalently,
$\theta_0(L)=2\arctan\frac12\approx0.927295$,
$\theta_0(R)=\pi$,
$\theta_1(L)=\pi$, and
$\theta_1(R)=2\arctan\frac1{\sqrt2}\approx1.230959$.
The last stage requires no rotation because the remaining
length is exactly one on both branches.

\paragraph{Step 3: Prepare the selector and implement the
branch-dependent rotations.}

We first prepare the branch selector. Since the two branches
have lengths five and three, the selector rotation is
$\phi=2\arccos\sqrt{\frac58}\approx1.318116$.
Therefore,
$R_y(\phi)\ket{0}_{q_3}
=\sqrt{\frac58}\ket{0}_{q_3}
+\sqrt{\frac38}\ket{1}_{q_3}$.
The squared amplitudes are precisely the relative branch
weights $L/M=5/8$ and $R/M=3/8$.

We next select the first block, of size four, using $q_2$.
For the lower branch the required angle is $\theta_0(L)$,
while for the upper branch it is $\pi$. We implement
these two angles with a baseline rotation followed by
a selector-controlled correction: first apply
$R_y(\theta_0(L))_{q_2}$ and then
$C_{q_3=1}\bigl(R_y(\Delta_0)_{q_2}\bigr)$, where
$\Delta_0=\theta_0(R)-\theta_0(L)\approx2.214297$.
Thus the total angle on the selector-zero branch is
$\theta_0(L)$, and on the selector-one branch it is
$\theta_0(L)+\Delta_0=\pi$.

After the selector rotation and these two block-selection
rotations, the state is
\begin{equation}\label{eq:worked-after-stage-zero}
  \ket{\psi_0}
  =\frac1{\sqrt2}\ket{0000}
   +\frac1{\sqrt8}\ket{0100}
   +\sqrt{\frac38}\ket{1100}.
\end{equation}

To see the amplitudes explicitly, consider the two branches
separately. On the lower branch, the selector amplitude is
$\sqrt{5/8}$ and
$R_y(\theta_0(L))\ket0
=\frac2{\sqrt5}\ket0+\frac1{\sqrt5}\ket1$.
The resulting amplitudes are therefore
$\sqrt{5/8}\,(2/\sqrt5)=1/\sqrt2$ and
$\sqrt{5/8}\,(1/\sqrt5)=1/\sqrt8$.
On the upper branch, the angle $\pi$ maps $\ket0$ to
$\ket1$, so the entire branch amplitude $\sqrt{3/8}$
is placed in the continuation.

We now select the next block, of size two, using $q_1$.
The required angles are $\theta_1(L)=\pi$ and
$\theta_1(R)=2\arccos\sqrt{2/3}$. A direct implementation
with two selector-dependent rotations would require a
two-qubit-controlled operation. Instead, we use the
paired-angle construction of Lemma~\ref{lem:paired-angle}.

Set $\Delta_1=\theta_1(R)-\theta_1(L)\approx-1.910633$
and
$\overline{\theta}_1
=\frac{\theta_1(L)+\theta_1(R)}2\approx2.186276$.
The target qubit $q_1$ is still in $\ket0$ before these
gates are applied. This is important: it allows the two
desired branch rotations to be implemented using three
singly controlled gates, in chronological order:
$C_{q_2=1}\bigl(R_y(-\Delta_1/2)_{q_1}\bigr)$,
$C_{q_3=1}(Z_{q_1})$, and
$C_{q_2=1}\bigl(R_y(\overline{\theta}_1)_{q_1}\bigr)$.
The two rotation angles are
$-\Delta_1/2\approx0.955317$ and
$\overline{\theta}_1\approx2.186276$.

Let us verify their action on the two branches. When
$q_3=0$ and $q_2=1$, the controlled-$Z$ gate is inactive.
The two rotations therefore combine to give
$R_y(\overline{\theta}_1)
R_y(-\Delta_1/2)\ket0
=R_y(\overline{\theta}_1-\Delta_1/2)\ket0
=R_y(\theta_1(L))\ket0=\ket1$.
Thus, the lower-branch component with $q_2=1$ is
transferred entirely to $q_1=1$.

When $q_3=1$ and $q_2=1$, the controlled-$Z$ gate is active.
Since $ZR_y(\alpha)\ket0=R_y(-\alpha)\ket0$, it reverses
the sign of the rotation angle on this input. Therefore,
$R_y(\overline{\theta}_1)
ZR_y(-\Delta_1/2)\ket0
=R_y(\overline{\theta}_1+\Delta_1/2)\ket0
=R_y(\theta_1(R))\ket0
=\sqrt{\frac23}\ket0+\frac1{\sqrt3}\ket1$.

If $q_2=0$, the two controlled rotations are inactive.
The target $q_1$ is still zero, so the controlled-$Z$ also
leaves the state unchanged. This verifies the required
action on every component of the state that can occur
in this example.

We can now see explicitly how the state changes. The
component $\frac1{\sqrt2}\ket{0000}$ is unchanged.
The lower-branch component $\frac1{\sqrt8}\ket{0100}$
becomes $\frac1{\sqrt8}\ket{0110}$. On the upper branch,
the component $\sqrt{3/8}\ket{1100}$ is split according
to the amplitudes $\sqrt{2/3}$ and $1/\sqrt3$:
\[
\sqrt{\frac38}\ket{1100}
\longmapsto
\sqrt{\frac38}
\left(
\sqrt{\frac23}\ket{1100}
+\frac1{\sqrt3}\ket{1110}
\right)
=\frac12\ket{1100}
+\frac1{\sqrt8}\ket{1110}.
\]

Combining these contributions gives
\begin{equation}\label{eq:worked-after-rotations}
  \ket{\psi_1}
  =\frac1{\sqrt2}\ket{0000}
   +\frac1{\sqrt8}\ket{0110}
   +\frac12\ket{1100}
   +\frac1{\sqrt8}\ket{1110}.
\end{equation}

The amplitudes now have a direct block interpretation.
The lower branch has selected a block of size four with
amplitude $1/\sqrt2=\sqrt{4/8}$ and a singleton block
with amplitude $1/\sqrt8$. The upper branch has selected
a block of size two with amplitude $1/2=\sqrt{2/8}$ and
a singleton block with amplitude $1/\sqrt8$. In each case,
the squared amplitude assigned to a block is its size
divided by the total interval size. The next step distributes
each block's amplitude uniformly among its computational-basis
states.

\paragraph{Step 4: Expand the selected dyadic blocks.}

We now apply the two common controlled-Hadamard gates,
in descending partition order:
$C_{q_1=0}(H_{q_0})$ and $C_{q_2=0}(H_{q_1})$.

The first gate applies a Hadamard to $q_0$ whenever
$q_1=0$. It therefore expands the selected blocks of
sizes four and two, while leaving the singleton blocks
unchanged. For example, the size-four block is represented
initially by $\frac1{\sqrt2}\ket{0000}$. The Hadamard
on $q_0$ gives
$\frac1{\sqrt2}\ket{0000}
\longmapsto\frac12(\ket{0000}+\ket{0001})$.
Similarly, the size-two block is represented by
$\frac12\ket{1100}$, and the same controlled-Hadamard gives
$\frac12\ket{1100}
\longmapsto\frac1{\sqrt8}(\ket{1100}+\ket{1101})$.

The singleton components, for which $q_1=1$, are
unaffected. The intermediate state is thus
\begin{align}
  \ket{\psi_2}={}&
  \frac12(\ket{0000}+\ket{0001})
  +\frac1{\sqrt8}
  (\ket{0110}+\ket{1100}+\ket{1101}+\ket{1110}).
  \label{eq:worked-after-first-expansion}
\end{align}

Next, the controlled-Hadamard on $q_1$ acts only when
$q_2=0$. This condition identifies the size-four block.
It therefore expands that block in the $q_1$ direction
and leaves the size-two and singleton blocks unchanged.
Explicitly,
$\frac12\ket{0000}\longmapsto
\frac1{\sqrt8}(\ket{0000}+\ket{0010})$
and
$\frac12\ket{0001}\longmapsto
\frac1{\sqrt8}(\ket{0001}+\ket{0011})$.

The resulting state is
\begin{align}
  \ket{\Phi}
  =\frac1{\sqrt8}(&\ket{0000}+\ket{0001}
       +\ket{0010}+\ket{0011}
       +\ket{0110}+\ket{1100}
       +\ket{1101}+\ket{1110}).
  \label{eq:worked-common-partition-state}
\end{align}

The order of the two controlled-Hadamards is essential.
We first expand the lowest-order free bit $q_0$, while the
higher-order marker $q_1$ is still fixed. We then expand
$q_1$ only in the size-four block, identified by $q_2=0$.
This way, the same two gates expand all selected blocks
without requiring additional branch controls.

Before the final remapping, the lower-register supports are
$\mathcal S_0=\{0,1,2,3,6\}$ and
$\mathcal S_1=\{4,5,6\}$ for the selector-zero and
selector-one branches, respectively. Every basis state
in these supports has amplitude $1/\sqrt8$. More generally,
a block of size $d$ initially carries amplitude $\sqrt{d/8}$;
its $s=\log_2d$ Hadamards distribute this amplitude equally
among its $d$ basis states, giving
$\frac{\sqrt{d/8}}{\sqrt d}=\frac1{\sqrt8}$
on each state.

\paragraph{Step 5: Apply the branch-dependent XOR masks.}

\begin{figure}[htbp]
    \centering
    \includegraphics[width=0.95\linewidth]
    {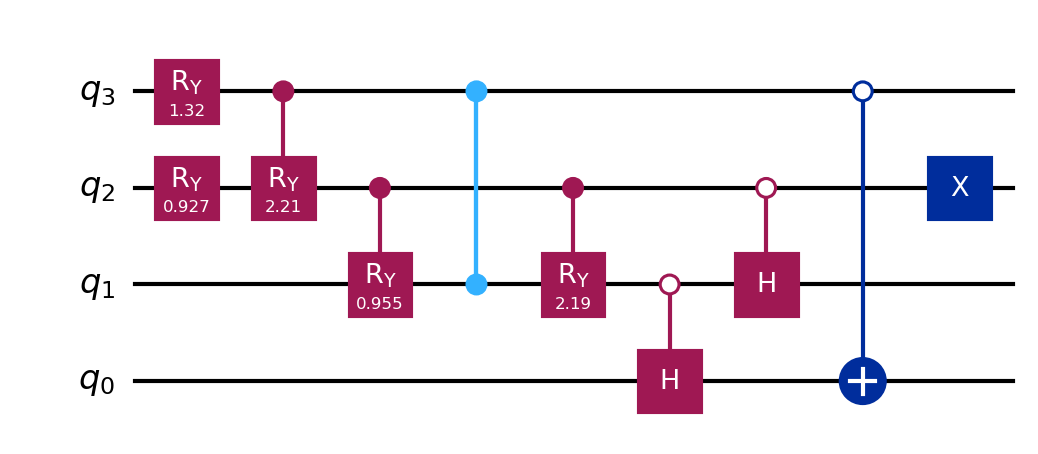}
    \caption{Qiskit circuit implementing the preparation of the
    uniform superposition over $[3,11)$ in a four-qubit register.}
    \label{fig:qiskit-interval-3-11}
\end{figure}

The state $\ket{\Phi}$ is already uniform, but its support
is not yet the desired interval. The common dyadic partition
has produced the same internal block structure for both
branches; we now use different XOR masks to move the two
supports to their required locations.

Since $L=5$ and $R=3$ are both odd, their 2-adic valuations
are zero. Equation~\eqref{eq:sigma-mask} therefore gives
$\sigma_L=(7\mathbin{\oplus}5)\mathbin{\land}(2^3-2)=2$
and
$\sigma_R=(7\mathbin{\oplus}3)\mathbin{\land}(2^3-2)=4$.
The corresponding branch masks are
$\tau_0=(2^3-1)\mathbin{\oplus}\sigma_L=5$ and
$\tau_1=\sigma_R=4$.

In binary, these are $\tau_0=101_2$ and $\tau_1=100_2$.
The two masks have the bit corresponding to $q_2$ in common.
We therefore implement that flip unconditionally, using
$X_{q_2}$. The lower-branch mask has one additional set bit,
corresponding to $q_0$, which is implemented by an $X$ gate
controlled on $q_3=0$. Thus, only two gates are needed:
$X_{q_2}$ and $C_{q_3=0}(X_{q_0})$.

The effective lower-register masks are therefore $5$ on
the selector-zero branch and $4$ on the selector-one branch.
We can check their action directly:
\[
\begin{aligned}
  \mathcal S_0\mathbin{\oplus}5
  &=\{0,1,2,3,6\}\mathbin{\oplus}5
  =\{3,4,5,6,7\},\\
  \mathcal S_1\mathbin{\oplus}4
  &=\{4,5,6\}\mathbin{\oplus}4
  =\{0,1,2\}.
\end{aligned}
\]

The selector-one branch also carries the contribution
$2^3=8$. Hence the two remapped branches occupy
$[3,8)$ and $[8,11)$, respectively. We have therefore
obtained the desired uniform superposition:
\[
  \ket{\psi_{3,8}^{(4)}}
  =\frac1{\sqrt8}\bigl(
  \ket3+\ket4+\ket5+\ket6+
  \ket7+\ket8+\ket9+\ket{10}\bigr).
\]

A Qiskit implementation of this four-qubit preparation is
shown in \Cref{fig:qiskit-interval-3-11}. It provides a
circuit-level realization of the construction described above.

\section{Conclusion}
\label{sec:conclusion}
We have constructed exact, ancilla-free circuits for preparing
uniform quantum states on arbitrary nonempty integer intervals.
The circuits use only one-qubit gates and singly controlled
one-qubit gates. In the logical model adopted here, the direct
construction has worst-case gate count at most $5n-9$ for $n\geq3$,
with exact maxima $1$ and $3$ for $n=1,2$. A lower bound of
$\Omega(n)$ establishes asymptotic optimality in register width.

In addition to this worst-case result, we have established an
instance-dependent gate-count bound
\[
G_{\mathrm{dir}}(a,M)
\leq4\lfloor\log_2M\rfloor+\epsilon(a,M),
\qquad M>1,
\]
where $\epsilon(a,M)$ is determined by the binary Hamming weights
of the common high-order prefix and, in the generic case, the
final branch-dependent masks. For singleton intervals, the exact
count is $\operatorname{wt}(a)$. This refinement separates the
logarithmic cost associated with the interval length from the
additional cost associated with its location and binary structure
within the register. In particular, intervals for which
$\epsilon(a,M)=O(\log M)$ admit $O(\log M)$ logical gate
complexity, even when embedded in a register of much larger width.

We have also given a reflected construction and a classical compiler
that selects the lower-count complete circuit while retaining the
same worst-case upper bound. The direct construction has logical depth at most $5n-9$
for $n\geq3$. This is
a linear upper bound; we do not establish a matching
lower bound on logical depth. Finally, we
established exact extensions to bitwise-XOR images of intervals
and to arithmetic progressions whose common difference is a
power of two. 

The common dyadic partition also admits an extension to
block-weighted nonuniform states, in which the probabilities
of individual blocks can be prescribed while retaining
uniform amplitudes within each block; this extension is
developed in Appendix~\ref{app:nonuniform-dyadic}.

\appendix
\section*{Appendix}
\section{Technical details of the reflected construction}
\label{app:reflected-suffix-details}

\subsection{A reflected circuit for an aligned suffix}
\label{subsec:reflected-suffix}

The reflected construction admits a particularly simple
single-branch specialization for aligned suffixes. Suppose
$2^k<M<2^{k+1}$ and put $K=k+1$. The desired state on the
$K$-qubit register is supported on $[2^K-M,2^K)$.

Write $M=2^{v_M}N_M$, where $N_M$ is odd, and let
$s_M=\operatorname{wt}(N_M)$. Denote the set-bit positions
of $N_M$ by $d_0>\cdots>d_{s_M-1}=0$, and put
$h_M=k-v_M+1$ and
$e_r=v_M+d_{s_M-1-r}$ for $0\leq r<s_M$.
Thus $e_0=v_M<\cdots<e_{s_M-1}=k$.
This is the single-length specialization of
Section~\ref{sec:reflected-construction}, with $S=M$
and $\lambda_0=k$. In particular, writing
$A_r=\sum_{s=0}^{r-1}2^{e_s}$, the reflected blocks are
$D_r=[2^K-M+A_r,\,2^K-M+A_r+2^{e_r})$.
They are pairwise disjoint and partition the desired suffix.
Initialize the register by applying $H$ to
$q_0,\ldots,q_{v_M-1}$ and $X$ to the qubits in
\[
 \mathcal I_0^{\mathrm{suffix}}
 =
 \{v_M,\ldots,k\}
 \setminus\{e_1,\ldots,e_{s_M-1}\}.
\]
For $0\leq r<s_M$, define
$V_r=\sum_{s=r}^{s_M-1}2^{e_s}$.
For each $r=0,\ldots,s_M-2$, use the rotation angle
\[
 \vartheta_r=
 2\arccos\sqrt{\frac{2^{e_r}}{V_r}}.
\]
Apply $R_y(\vartheta_0)$ to $q_{e_1}$ at the first
stage, and, for $r\geq1$, apply
$R_y(\vartheta_r)$ to $q_{e_{r+1}}$ controlled on
$q_{e_r}=1$. After all selection rotations, apply
$\C{q_{e_{r+1}}=1}{(XHX)_{q_t}}$ for
$r=0,\ldots,s_M-2$ and
$t=e_r,\ldots,e_{r+1}-1$.

All blocks are selected in this specialization. At stage $r$,
the rotation assigns squared norm $2^{e_r}/M$ to $D_r$,
and the remaining squared norm is carried to the next stage.
The expansion operations put precisely the lower $e_r$
qubits into $\ket+$ on $D_r$. Therefore, every basis
state in every selected block has amplitude $1/\sqrt M$.
The final residual component is the last block
$D_{s_M-1}$ and requires no additional rotation or
expansion. Since the blocks partition the suffix, the
output is
\[
 X^{\otimes K}\ket{\psi_M^{(K)}}
 =
 \frac{1}{\sqrt M}
 \sum_{x=2^K-M}^{2^K-1}\ket{x}.
\]

The circuit has $h_M-s_M+1$ initial bit flips,
$v_M$ Hadamard gates, $s_M-1$ selection rotations,
and $h_M-1$ controlled $XHX$ gates. Its logical gate
count is therefore the reflected suffix count
$C_0=K+h_M-1$, obtained in
\eqref{eq:reflected-C0}. The common high-order prefix
contributes an additional $\operatorname{wt}(Q)$ gates.
This is the aligned-suffix circuit selected by the
adaptive compiler in \Cref{alg:adaptive}.

\section{Additional complexity proofs}
\label{app:additional-complexity-proofs}

\subsection{Proof of the instance-dependent gate bound}

\begin{proof}
We consider the generic and aligned cases separately.

\medskip
\noindent
\emph{Generic intervals.}
Suppose $0<L,R<2^k$, and write
$S=L\mathbin{\lor}R=\sum_{j=0}^{h-1}2^{\lambda_j}$,
where $\lambda_0>\cdots>\lambda_{h-1}=v$.
Put $\lambda_0=\lfloor\log_2 S\rfloor$ and
$p=\lfloor\log_2 M\rfloor$. Since $S\leq L+R=M$,
we have $\lambda_0\leq p$, while
$h\leq\lambda_0+1$.

First suppose $h\geq2$. By
\eqref{eq:partition-count-upper}, the direct construction
satisfies
\[
G_{\mathrm{dir}}(a,M)
\leq \lambda_0+h+\delta_0+2r
+\operatorname{wt}(\widehat{\tau}_L
\mathbin{\lor}\widehat{\tau}_R)
+\operatorname{wt}(Q),
\]
where $\delta_0\leq1$ and $r\leq h-2$.
The terms independent of the final masks satisfy
$\lambda_0+h+\delta_0+2r
\leq\lambda_0+h+1+2(h-2)
=\lambda_0+3h-3
\leq4\lambda_0\leq4p$.
By the definition of $\epsilon(a,M)$, it follows that
$G_{\mathrm{dir}}(a,M)\leq4p+\epsilon(a,M)$.

It remains to consider $h=1$. Since $L$ and $R$ are
positive submasks of $S$, this case implies
$L=R=S=2^v$ and $M=2^{v+1}$.
The common-partition circuit uses $v$ initial Hadamards
and one selector rotation. Moreover,
$\sigma_L=\sigma_R=0$, so that
$\widehat{\tau}_L=2^k-2^v$ and
$\widehat{\tau}_R=0$. The final branch-mask conversion
therefore requires $k-v$ gates. Including the common
high-order prefix, the total is
$G_{\mathrm{dir}}(a,M)
=v+1+(k-v)+\operatorname{wt}(Q)
=\lfloor\log_2M\rfloor+\epsilon(a,M)$.
This is no greater than
$4\lfloor\log_2M\rfloor+\epsilon(a,M)$, completing
the generic case.

\medskip
\noindent
\emph{Aligned prefix.}
Suppose $L=2^k$ and $0<R<2^k$. Then
$M=2^k+R$ and $p=k$.
Write $M=2^{v_M}N_M$, where $N_M$ is odd, and put
$s_M=\operatorname{wt}(N_M)$.
The prefix-state circuit uses $s_M-1$ initial bit flips,
$v_M$ Hadamards, $s_M-1$ rotations, and
$k-v_M$ controlled Hadamards. Its count on the lower
$k+1$ qubits is therefore $k+2s_M-2$.

Since $s_M\leq k-v_M+1$, this count satisfies
$k+2s_M-2\leq3k-2v_M\leq3k$.
Adding the $\operatorname{wt}(Q)$ gates required to set
the common high-order prefix gives
$G_{\mathrm{dir}}(a,M)\leq
3p+\operatorname{wt}(Q)
\leq4p+\epsilon(a,M)$.

\medskip
\noindent
\emph{Aligned suffix.}
Suppose $R=2^k$ and $0<L<2^k$. Again,
$M=2^k+L$ and $p=k$.
Write $M=2^{v_M}N_M$, where $N_M$ is odd, and put
$s_M=\operatorname{wt}(N_M)$ and
$h_M=k-v_M+1$.

The direct complement-adapted suffix circuit has lower-register
count $C_1=K+h_M+s_M-3$, where $K=k+1$.
Since $s_M\leq h_M$, we obtain
$C_1=2k-v_M+s_M-1\leq3k-2v_M\leq3k$.

The reflected suffix circuit selected by the adaptive compiler
has count $C_0=K+h_M-1$. By
\eqref{eq:reflected-count-difference}, we have
$C_0\leq C_1$. Thus both the direct and selected reflected
realizations satisfy the same instance-dependent estimate.
Including the common high-order prefix, we obtain
$G_{\mathrm{dir}}(a,M)\leq3p+\operatorname{wt}(Q)$
and $G_{\mathrm{ad}}(a,M)\leq3p+\operatorname{wt}(Q)$.
Since $\epsilon(a,M)=\operatorname{wt}(Q)$ in this case,
both counts are bounded by $4p+\epsilon(a,M)$.

\medskip
\noindent
\emph{Full parent block.}
If $L=R=2^k$, the interval is the entire parent block.
It requires $k+1$ Hadamards on the lower register and
$\operatorname{wt}(Q)$ gates for the common high-order
prefix. Since $M=2^{k+1}$, the total count is
$G_{\mathrm{dir}}(a,M)
=\lfloor\log_2M\rfloor+\operatorname{wt}(Q)
\leq4\lfloor\log_2M\rfloor+\epsilon(a,M)$.
The adaptive compiler uses this same specialized circuit.

\medskip
\noindent
\emph{Singleton intervals.}
For $M=1$, the target is the computational-basis state
$\ket a$. Applying one Pauli-$X$ gate at each set-bit
position of $a$ gives the exact count
$G_{\mathrm{dir}}(a,1)=\operatorname{wt}(a)$.

Finally, for generic intervals the adaptive compiler
selects the smaller of the direct and reflected circuit
counts. For aligned suffixes it selects the reflected
circuit, whose count is no greater than that of the direct
complement-adapted circuit. For aligned prefixes, full
parent blocks, and singleton intervals, it retains the
corresponding direct specialized circuit. Therefore, for
every nonempty interval,
$G_{\mathrm{ad}}(a,M)\leq G_{\mathrm{dir}}(a,M)$.

It follows that the adaptive compiler satisfies the same
instance-dependent bound:
$G_{\mathrm{ad}}(a,M)
=O(\log_2M+\epsilon(a,M))$ for $M>1$.
For $M=1$, its exact count is $\operatorname{wt}(a)$.
\end{proof}

\subsection{Proof of the logical-depth upper bound}

\begin{proof}
The first term in \eqref{eq:generic-depth} accounts for
the single initialization layer. The next two layers
accommodate the at most two stage-zero rotations.
Each of the $h-2$ subsequent stages uses at most three
gates in sequence by Lemma~\ref{lem:paired-angle}.

The number of controlled-Hadamard expansion gates is
$\sum_{j=0}^{h-2}(\lambda_j-\lambda_{j+1})
=\lambda_0-v$. Placing these gates in separate layers
therefore gives a valid, though not necessarily optimal,
schedule. The final masks require at most $k$ layers,
as established above. This proves
\eqref{eq:generic-depth}.

Since $\lambda_0-v\leq k-1$, $h\leq k$, and
$k\leq n-1$, we obtain
$D(a,M)\leq(k-1)+3k+k-3=5k-4\leq5n-9$.

For aligned-prefix, aligned-suffix, and one-block intervals,
the depth is no greater than the corresponding logical
gate count. The bounds established in the proof of
Theorem~\ref{thm:interval-upper} are also at most $5n-9$
for $n\geq3$. Singleton intervals are prepared by parallel
Pauli-$X$ gates and therefore have depth at most one.
The two small-register cases follow by direct inspection.

This establishes the stated depth bound for the direct
construction. The adaptive compiler also inherits the
global depth bound: by
Proposition~\ref{prop:instance-dependent-bound},
$G_{\mathrm{ad}}(a,M)\leq G_{\mathrm{dir}}(a,M)$, while
any returned circuit can be scheduled serially with one
gate per layer. Hence
\[
  D_{\mathrm{ad},1}\leq1,\qquad
  D_{\mathrm{ad},2}\leq3,\qquad
  D_{\mathrm{ad},n}\leq5n-9
  \quad(n\geq3).
\]
This worst-case guarantee does not imply that minimizing
logical gate count minimizes logical depth for an
individual interval.
\end{proof}

\section{Additional interval identities}
\label{app:dyadic-parent-complement}

\begin{corollary}[Complement within the dyadic parent block]
\label{cor:dyadic-parent-complement}
Let $I=[a,a+M)$ with $M>1$ and midpoint-split parameters
$C,k,L,R$, and let
$  P=[C,C+2^{k+1})$
be its dyadic parent block. Suppose $M<2^{k+1}$, so
that the complement $P\setminus I$ is nonempty. Define
  $  J=[C+R,C+2^{k+1}-L)$.
Then
   $ P\setminus I
    =
    \{x\mathbin{\oplus}2^k:x\in J\}$.
Therefore, the normalized uniform state on $P\setminus I$
can be prepared by preparing the interval state on $J$ and
then applying $X$ to $q_k$. This adds one logical gate to
the interval-preparation circuit for $J$.
\end{corollary}

\begin{proof}
Put $H=2^k$. By the midpoint-split definitions,
  $  a=C+H-L$ and $a+M=C+H+R$.
The interval $J$ splits at the midpoint $C+H$ into
   $ [C+R,C+H)$ and
    $[C+H,C+2H-L)$,
where either part may be empty.
Since $C$ is divisible by $2H$, bitwise XOR by $H$
exchanges the two halves of the parent block. Thus these
two subintervals are mapped to
    $[C+H+R,C+2H)$ and 
   $ [C,C+H-L)$,
respectively. Their union is precisely $P\setminus I$.
Since XOR preserves amplitudes, the state-preparation
claim follows.
\end{proof}
For example, if $I=[3,13)$ and $P=[0,16)$, then
$J=[5,11)$. Applying $X$ to $q_3$ maps this interval to
$[0,3)\cup[13,16)$,
which is the complement of $[3,13)$ in $[0,16)$.

\section{Nonuniform superpositions over the common dyadic partition}
\label{app:nonuniform-dyadic}

The common dyadic-block construction also extends naturally to
nonuniform state preparation. As shown in our earlier work
\cite{shukla2024efficient}, nonuniform weights on a dyadic
decomposition can be incorporated by modifying the rotation
angles used during block selection. The same principle applies
to the common partition of an arbitrary interval. In particular,
the block partition, controlled-Hadamard expansion, and
branch-dependent XOR masks remain unchanged.

\subsection{Weighted preparation over the common dyadic partition}

%We use the notation of Lemma~\ref{lem:common-partition}.
We consider the generic midpoint split $1\leq L,R<2^k$
and use the notation of Lemma~\ref{lem:common-partition}.
Let $J_0=\{j:b_j(L)=1\}$ and
$J_1=\{j:b_j(R)=1\}$ denote the block-index sets of the
lower and upper branches, respectively. Write
$\tau_0=\tau_L$ and $\tau_1=\tau_R$, where
$\tau_L$ and $\tau_R$ are the final branch-dependent XOR
masks defined in \eqref{eq:final-branch-masks}.
For $s\in\{0,1\}$ and $j\in J_s$, the normalized masked
block state is
$\ket{\mathcal D_{s,j}}
=2^{-\lambda_j/2}\sum_{x\in B_j}
\ket{x\mathbin{\oplus}\tau_s}$.
The common high-order prefix $Q$ is suppressed in this notation.

Assign nonnegative probabilities $p_{s,j}$ to the occupied
blocks, with $p_{s,j}=0$ for $j\notin J_s$ and
$\sum_{s=0}^{1}\sum_{j=0}^{h-1}p_{s,j}=1$.
The target state is
\begin{equation}
\label{eq:app-nonuniform-common-partition}
\ket{\Omega}
=
\sum_{s=0}^{1}\sum_{j\in J_s}
\sqrt{p_{s,j}}\,
\ket{s}_{q_k}\ket{\mathcal D_{s,j}}.
\end{equation}

The required modification concerns only the rotation angles.
First, put $P_s=\sum_{j=0}^{h-1}p_{s,j}$.
Only branches with $P_s>0$ are included in the preparation.
If $P_s=0$, then $p_{s,j}=0$ for every $j$, and the
corresponding branch is omitted entirely from the
block-selection procedure.
The selector qubit is prepared by applying
$R_y(2\arccos\sqrt{P_0})$ to $q_k$, initially in
$\ket0$, producing
$\sqrt{P_0}\ket0_{q_k}+\sqrt{P_1}\ket1_{q_k}$.
The cases $P_0=0$ and $P_1=0$ are included: the selector
is then prepared directly in $\ket1$ and $\ket0$,
respectively.

For each branch, define the residual probability
$T_{s,j}=\sum_{\ell=j}^{h-1}p_{s,\ell}$.
At stage $j=0,\ldots,h-2$, whenever $T_{s,j}>0$,
replace the uniform block-selection angle by
\[
\vartheta_{s,j}
=
2\arccos\sqrt{\frac{p_{s,j}}{T_{s,j}}}.
\]
Thus, a block with zero weight gives angle $\pi$, whereas
a block containing the entire residual probability gives
angle zero. A terminated branch requires no further operation,
and no rotation is needed at the final block position.

At the first selection stage, $j=0$, if both branches are
active, the two branch-dependent angles are implemented using
a selector-controlled correction, as in the direct construction.
If one branch has zero total probability, the selector is
deterministic and the active branch's rotation is applied as
a single one-qubit rotation, without a selector-controlled
correction.
At each subsequent stage, $j\geq1$, if both branches remain
active, equal angles require only a single controlled rotation,
whereas unequal angles are implemented using
Lemma~\ref{lem:paired-angle}. This includes the case in which
one of the two angles is zero.
If exactly one branch remains active, its rotation is
implemented by a single controlled rotation on
$q_{\lambda_j}$, controlled on
$q_{\lambda_{j-1}}=1$. In the terminated branch, the
preceding continuation qubit is in $\ket0$, so this operation
leaves that branch unchanged. If both branches have terminated,
the stage is omitted.
The controlled-Hadamard expansion and the final
branch-dependent XOR masks are exactly those of the
uniform construction.

To verify the probability assignment, observe that
$\cos^2(\vartheta_{s,j}/2)=p_{s,j}/T_{s,j}$.
Therefore, a rotation assigns squared norm $p_{s,j}$
to the current block and leaves $T_{s,j+1}$ in the
continuation component. Successive application assigns
the prescribed probability to every occupied block.
The unchanged controlled-Hadamard gates expand each
selected block uniformly, while the XOR masks place it
at the required location. Hence the resulting state is
exactly \eqref{eq:app-nonuniform-common-partition}.
The construction remains ancilla-free and uses only
one-qubit and singly controlled one-qubit gates.

\subsection{Recovery of the uniform interval state}

The uniform interval state is recovered by assigning
each occupied block a probability proportional to its
size, namely $p_{s,j}=2^{\lambda_j}/M$ for $j\in J_s$,
and zero otherwise. Since the occupied blocks partition
the interval and have total size $M$, these probabilities
sum to one. Thus, the original uniform interval-preparation
algorithm is a special case of the block-weighted construction.

\subsection{Worked example: a nonuniform state on
\texorpdfstring{$[3,11)$}{[3,11)}}
\label{app:nonuniform-example}

Consider the interval $[3,11)$, for which $a=3$, $M=8$,
and $k=3$. The midpoint split gives $L=5$, $R=3$, and
$D=8$. The common partition is determined by
$S=L\mathbin{\lor}R=5\mathbin{\lor}3=7
=2^2+2^1+2^0$. Thus, the canonical blocks are
$B_0=[0,4)$, $B_1=[4,6)$, and $B_2=[6,7)$.
The lower and upper branches use $J_0=\{0,2\}$ and
$J_1=\{1,2\}$, respectively, with branch masks
$\tau_L=5$ and $\tau_R=4$.

Assign the nonzero block probabilities
$p_{0,0}=1/4$, $p_{0,2}=1/4$,
$p_{1,1}=7/20$, and $p_{1,2}=3/20$;
all other weights are zero. The two branch probabilities
are $P_0=P_1=1/2$, so the selector is prepared using
$R_y(\pi/2)$.

The residual probabilities are
$(T_{0,0},T_{0,1},T_{0,2})=(1/2,1/4,1/4)$
and
$(T_{1,0},T_{1,1},T_{1,2})=(1/2,1/2,3/20)$.
The required block-selection angles are therefore
$\vartheta_{0,0}=\pi/2$, $\vartheta_{0,1}=\pi$,
$\vartheta_{1,0}=\pi$, and
$\vartheta_{1,1}=2\arccos\sqrt{7/10}$.
No rotation is required at the final block position.
The branch-dependent rotations are implemented using
the selector-controlled correction at the first stage
and the three-gate paired-angle identity at the second.

After the controlled-Hadamard expansion and the
branch-dependent XOR masks, the resulting state is
\begin{equation}
\label{eq:app-nonuniform-example-state}
%\begin{aligned}
\ket{\Omega}
=\frac12\ket{3}
+\frac14\sum_{x=4}^{7}\ket{x}%\\
+\sqrt{\frac7{40}}\bigl(\ket8+\ket9\bigr)
+\sqrt{\frac3{20}}\ket{10}.
%\end{aligned}
\end{equation}
The corresponding measurement probabilities are $1/4$
at $x=3$, $1/16$ at each of $x=4,\ldots,7$,
$7/40$ at each of $x=8,9$, and $3/20$ at $x=10$.
They sum to one and are constant within each physical
dyadic block, but differ between blocks.

\begin{figure}[ht]
\centering
\begin{tikzpicture}
\begin{axis}[
    width=0.85\linewidth,
    height=6cm,
    ybar,
    bar width=12pt,
    xlabel={Computational-basis label $x$},
    ylabel={Measurement probability},
    xtick={3,4,5,6,7,8,9,10},
    ymin=0,
    ymax=0.31,
    ytick={0,0.05,0.10,0.15,0.20,0.25,0.30},
    grid=major,
    enlarge x limits=0.08,
]
\addplot coordinates {
    (3,0.25)
    (4,0.0625)
    (5,0.0625)
    (6,0.0625)
    (7,0.0625)
    (8,0.175)
    (9,0.175)
    (10,0.15)
};

% Boundaries of the four physical dyadic blocks
\draw[dashed] (axis cs:3.5,0) -- (axis cs:3.5,0.30);
\draw[dashed] (axis cs:7.5,0) -- (axis cs:7.5,0.30);
\draw[dashed] (axis cs:9.5,0) -- (axis cs:9.5,0.30);

\node[font=\scriptsize] at (axis cs:3,0.275)
    {$1/4$};
\node[font=\scriptsize] at (axis cs:5.5,0.09)
    {$1/16$};
\node[font=\scriptsize] at (axis cs:8.5,0.205)
    {$7/40$};
\node[font=\scriptsize] at (axis cs:10,0.18)
    {$3/20$};
\end{axis}
\end{tikzpicture}
\caption{Measurement probabilities of the nonuniform
state supported on $[3,11)$. Dashed lines separate
the physical dyadic blocks, within each of which
the probability is constant.}
\label{fig:app-nonuniform-interval-histogram}
\end{figure}
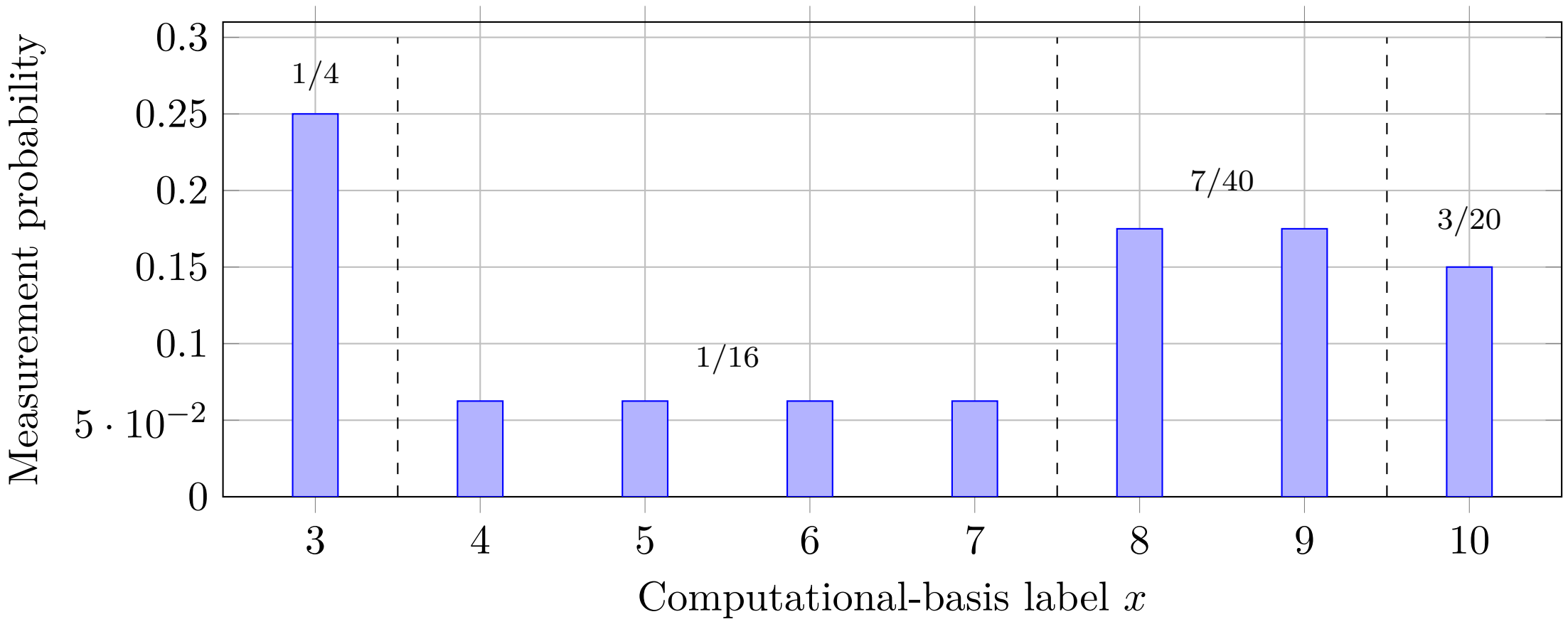

The example illustrates that the common dyadic partition
is not restricted to uniform interval states: changing the
block-selection rotation angles prescribes the probability
assigned to each occupied block while retaining uniform
amplitudes within that block. The construction assigns
nonnegative real amplitudes to the normalized block states;
arbitrary relative complex phases between different blocks
would additionally require phase operations.

\section{Further interval extensions}
\label{app:two-disjoint-intervals}

\begin{remark}[Preparation of two translated copies of an interval]
\label{rem:two-intervals}

The interval-preparation construction also gives an immediate
extension to two translated copies of the same interval.
Suppose $0\leq a<a+M\leq 2^k$, and let
\[
\ket{\psi_{a,M}^{(k)}}
=\frac{1}{\sqrt M}\sum_{x=a}^{a+M-1}\ket{x}
\]
be the corresponding $k$-qubit interval state.
Introduce one additional qubit $q_k$ immediately above
the interval register and prepare it in $\ket{+}$.
The resulting state is
\[
\begin{aligned}
\ket{\Psi}
=\ket{+}_{q_k}\otimes\ket{\psi_{a,M}^{(k)}} =\frac{1}{\sqrt{2M}}\left(
\sum_{x=a}^{a+M-1}\ket{0}\ket{x}
+\sum_{x=a}^{a+M-1}\ket{1}\ket{x}
\right).
\end{aligned}
\]
Interpreting the two registers as a $(k+1)$-qubit
binary register, this is precisely the uniform
superposition over
\[
[a,a+M)\ \cup\
[2^k+a,\,2^k+a+M).
\]

The two intervals have the same length and are
translates of one another by $2^k$. They are disjoint,
and the resulting state is prepared by the original
interval-preparation circuit together with a single
Hadamard gate. More generally, the additional qubit
can be placed between a common high-order prefix and
the $k$-qubit interval register, producing the
corresponding two translated intervals within that
prefix block.

This construction concerns two translated copies of
a common interval; it does not provide a preparation
method for a general union of two arbitrary disjoint
intervals. The additional qubit is part of the data
register and does not serve as workspace.
\end{remark}
% \clearpage

\section*{Declaration on the Use of Generative AI}

Generative AI tools were used to assist with language
editing and improving the presentation of the manuscript,
as well as in developing simulation code in support of
the research. The authors reviewed the AI-assisted
material and take full responsibility for the content
of the manuscript, including the correctness of all
mathematical results and reported computational findings.

\bibliographystyle{unsrt}
%\bibliography{references}

\begin{thebibliography}{10}

\bibitem{deutsch1992rapid}
David Deutsch and Richard Jozsa.
\newblock Rapid solution of problems by quantum computation.
\newblock {\em Proceedings of the Royal Society of London. Series A: Mathematical and Physical Sciences}, 439(1907):553--558, 1992.

\bibitem{bernstein1993quantum}
Ethan Bernstein and Umesh Vazirani.
\newblock Quantum complexity theory.
\newblock In {\em Proceedings of the twenty-fifth annual ACM symposium on Theory of computing}, pages 11--20, 1993.

\bibitem{shukla2023generalizationBV}
Alok Shukla and Prakash Vedula.
\newblock A generalization of {B}ernstein--{V}azirani algorithm with multiple secret keys and a probabilistic oracle.
\newblock {\em Quantum Information Processing}, 22(6):244, 2023.

\bibitem{simon1997power}
Daniel~R Simon.
\newblock On the power of quantum computation.
\newblock {\em SIAM journal on computing}, 26(5):1474--1483, 1997.

\bibitem{grover1996fast}
Lov~K Grover.
\newblock A fast quantum mechanical algorithm for database search.
\newblock In {\em Proceedings of the Twenty-eighth Annual ACM Symposium on Theory of Computing}, pages 212--219. ACM, 1996.

\bibitem{shukla2025efficientGrover}
Alok Shukla and Prakash Vedula.
\newblock An efficient implementation of a quantum search algorithm for arbitrary {N}.
\newblock {\em The European Physical Journal Plus}, 140(6):575, 2025.

\bibitem{nielsen2002quantum}
Michael~A. Nielsen and Isaac Chuang.
\newblock {\em Quantum {C}omputation and {Q}uantum {I}nformation}.
\newblock Cambridge University Press, 2000.

\bibitem{kitaev1995quantum}
A~Yu Kitaev.
\newblock Quantum measurements and the abelian stabilizer problem.
\newblock {\em arXiv preprint quant-ph/9511026}, 1995.

\bibitem{dobvsivcek2007arbitrary}
Miroslav Dob{\v{s}}{\'\i}{\v{c}}ek, G{\"o}ran Johansson, Vitaly Shumeiko, and G{\"o}ran Wendin.
\newblock Arbitrary accuracy iterative quantum phase estimation algorithm using a single ancillary qubit: A two-qubit benchmark.
\newblock {\em Physical Review A—Atomic, Molecular, and Optical Physics}, 76(3):030306, 2007.

\bibitem{shukla2025towardsQPE}
Alok Shukla and Prakash Vedula.
\newblock Toward practical quantum phase estimation: A modular, scalable, and adaptive approach.
\newblock {\em Advanced Quantum Technologies}, 9(3):e00683, 2026.

\bibitem{brassard2002quantum}
Gilles Brassard, Peter Hoyer, Michele Mosca, and Alain Tapp.
\newblock Quantum amplitude amplification and estimation.
\newblock {\em Contemporary Mathematics}, 305:53--74, 2002.

\bibitem{giurgica2022low}
Tudor Giurgica-Tiron, Iordanis Kerenidis, Farrokh Labib, Anupam Prakash, and William Zeng.
\newblock Low depth algorithms for quantum amplitude estimation.
\newblock {\em Quantum}, 6:745, 2022.

\bibitem{shukla2025modularQAE}
Alok Shukla and Prakash Vedula.
\newblock Modular quantum amplitude estimation: A scalable and adaptive framework.
\newblock {\em arXiv preprint arXiv:2508.05805}, 2025.

\bibitem{shor1999polynomial}
Peter~W Shor.
\newblock Polynomial-time algorithms for prime factorization and discrete logarithms on a quantum computer.
\newblock {\em SIAM review}, 41(2):303--332, 1999.

\bibitem{shukla2025modularShor}
Alok Shukla and Prakash Vedula.
\newblock A modular, adaptive, and scalable quantum factoring algorithm.
\newblock {\em The European Physical Journal Plus}, 141(4):474, 2026.

\bibitem{shukla2019trajectory}
Alok Shukla and Prakash Vedula.
\newblock Trajectory optimization using quantum computing.
\newblock {\em Journal of Global Optimization}, 75(1):199--225, 2019.

\bibitem{harrow2009quantum}
Aram~W Harrow, Avinatan Hassidim, and Seth Lloyd.
\newblock Quantum algorithm for linear systems of equations.
\newblock {\em Physical Review Letters}, 103(15):150502, 2009.

\bibitem{childs2020quantum}
Andrew~M Childs and Jin-Peng Liu.
\newblock Quantum spectral methods for differential equations.
\newblock {\em Communications in Mathematical Physics}, 375(2):1427--1457, 2020.

\bibitem{berry2014high}
Dominic~W Berry.
\newblock High-order quantum algorithm for solving linear differential equations.
\newblock {\em Journal of Physics A: Mathematical and Theoretical}, 47(10):105301, 2014.

\bibitem{shukla2023hybriddiffeq}
Alok Shukla and Prakash Vedula.
\newblock A hybrid classical-quantum algorithm for solution of nonlinear ordinary differential equations.
\newblock {\em Applied Mathematics and Computation}, 442:127708, 2023.

\bibitem{shukla2023hybridimageprocessing}
Alok Shukla and Prakash Vedula.
\newblock A hybrid classical-quantum algorithm for digital image processing.
\newblock {\em Quantum Information Processing}, 22(1):1--19, 2023.

\bibitem{shukla2023quantumDSP}
Alok Shukla and Prakash Vedula.
\newblock A quantum approach for digital signal processing.
\newblock {\em The European Physical Journal Plus}, 138(12):1--24, 2023.

\bibitem{rohida2024hybridPolar}
Mohit Rohida, Alok Shukla, and Prakash Vedula.
\newblock Hybrid classical-quantum image processing via polar walsh basis functions.
\newblock {\em Quantum Machine Intelligence}, 6(2):72, 2024.

\bibitem{rohida2025quantumEdgeDetect}
Mohit Rohida, Alok Shukla, and Prakash Vedula.
\newblock Quantum algorithm for edge detection in digital grayscale images.
\newblock {\em arXiv preprint arXiv:2507.06642}, 2025.

\bibitem{sandesara2024quantumOptimalcontrol}
Hirmay Sandesara, Alok Shukla, and Prakash Vedula.
\newblock A quantum approach for optimal control.
\newblock {\em Quantum Inf Process 24, 95 (2025)}, 2024.

\bibitem{biamonte2017quantum}
Jacob Biamonte, Peter Wittek, Nicola Pancotti, Patrick Rebentrost, Nathan Wiebe, and Seth Lloyd.
\newblock Quantum machine learning.
\newblock {\em Nature}, 549(7671):195--202, 2017.

\bibitem{shimada2020quantum}
N.~H. Shimada and T.~Hachisuka.
\newblock Quantum coin method for numerical integration.
\newblock {\em Computer Graphics Forum}, 39(6):243--257, 2020.

\bibitem{abrams1999fast}
Daniel~S Abrams and Colin~P Williams.
\newblock Fast quantum algorithms for numerical integrals and stochastic processes.
\newblock {\em arXiv preprint quant-ph/9908083}, 1999.

\bibitem{shukla2025efficientPartialSum}
Alok Shukla and Prakash Vedula.
\newblock Efficient quantum algorithm for weighted partial sums and numerical integration.
\newblock {\em Advanced Quantum Technologies}, page e2500084, 2025.

\bibitem{grover1997quantum}
Lov~K Grover.
\newblock Quantum mechanics helps in searching for a needle in a haystack.
\newblock {\em Physical Review Letters}, 79(2):325, 1997.

\bibitem{shukla2024efficient}
Alok Shukla and Prakash Vedula.
\newblock An efficient quantum algorithm for preparation of uniform quantum superposition states.
\newblock {\em Quantum Information Processing}, 23(2):38, 2024.

\bibitem{QiskitUniformSuperpositionGate}
{IBM Quantum}.
\newblock {Qiskit SDK} 1.2 release notes, entry ``{UniformSuperpositionGate}''.
\newblock \url{https://qiskit.qotlabs.org/docs/api/qiskit/release-notes/1.2}, 2024.
\newblock Accessed 6 November 2026.

\bibitem{CirqUniformSuperpositionGate}
{Google Quantum AI}.
\newblock {Cirq API} reference, entry {\texttt{cirq.uniformsuperpositiongate}}.
\newblock \url{https://quantumai.google/reference/python/cirq/UniformSuperpositionGate}.
\newblock Accessed 6 November 2026.

\bibitem{mottonen2004transformation}
Mikko M{\"o}tt{\"o}nen, Juha~J. Vartiainen, Ville Bergholm, and Martti~M. Salomaa.
\newblock Transformation of quantum states using uniformly controlled rotations.
\newblock {\em Quantum Info. Comput.}, 5(6):467–473, September 2005.

\bibitem{ShendeBullockMarkov2006}
Vivek~V Shende, Stephen~S Bullock, and Igor~L Markov.
\newblock Synthesis of quantum logic circuits.
\newblock In {\em Proceedings of the 2005 Asia and South Pacific Design Automation Conference}, pages 272--275, 2005.

\bibitem{PleschBrukner2011}
Martin Plesch and {\v{C}}aslav Brukner.
\newblock Quantum-state preparation with universal gate decompositions.
\newblock {\em Physical Review A—Atomic, Molecular, and Optical Physics}, 83(3):032302, 2011.

\bibitem{GroverRudolph2002}
Lov Grover and Terry Rudolph.
\newblock Creating superpositions that correspond to efficiently integrable probability distributions.
\newblock {\em arXiv preprint quant-ph/0208112}, 2002.

\bibitem{McArdleGilyenBerta2026}
Sam McArdle, Andr{\'a}s Gily{\'e}n, and Mario Berta.
\newblock Quantum state preparation without coherent arithmetic.
\newblock {\em Physical Review Letters}, 136(24):240603, 2026.

\bibitem{babbush2018encoding}
Ryan Babbush, Craig Gidney, Dominic~W Berry, Nathan Wiebe, Jarrod McClean, Alexandru Paler, Austin Fowler, and Hartmut Neven.
\newblock Encoding electronic spectra in quantum circuits with linear t complexity.
\newblock {\em Physical Review X}, 8(4):041015, 2018.

\bibitem{Suresh2026PyEncode}
Krishnan Suresh and Sanjay Suresh.
\newblock Pyencode: an open-source library for structured quantum state preparation.
\newblock {\em Quantum Science and Technology}, 11(4):045032, 2026.

\bibitem{Draper2000}
Thomas~G Draper.
\newblock Addition on a quantum computer.
\newblock {\em arXiv preprint quant-ph/0008033}, 2000.

\bibitem{MozafariRienerSoekenDeMicheli2021}
Fereshte Mozafari, Heinz Riener, Mathias Soeken, and Giovanni De~Micheli.
\newblock Efficient boolean methods for preparing uniform quantum states.
\newblock {\em IEEE Transactions on Quantum Engineering}, 2:1--12, 2021.

\bibitem{MozafariDeMicheliYang2022}
Fereshte Mozafari, Giovanni De~Micheli, and Yuxiang Yang.
\newblock Efficient deterministic preparation of quantum states using decision diagrams.
\newblock {\em Physical Review A}, 106(2):022617, 2022.

\end{thebibliography}

\end{document}